\documentclass{article}
\usepackage{authblk} 
\usepackage{graphicx}
\usepackage[utf8]{inputenc}
\usepackage[T1]{fontenc}
\usepackage{bm}
\usepackage[a4paper, total={7in, 8in}]{geometry}
\usepackage{multicol}
\usepackage{float}
\usepackage{blindtext}
\usepackage{siunitx}
\usepackage[superscript, biblabel, nomove]{cite}
\usepackage{xcolor}
\usepackage{amsmath}
\usepackage{amsfonts} 
\usepackage{mathtools}
\usepackage{nccmath}
\usepackage{verbatim}
\usepackage{longtable}
\usepackage{booktabs}
\usepackage[normalem]{ulem}
\usepackage{textgreek}

\newcommand{\tabhead}[1]{\textbf{#1}}

\title{Spectral tuning of hyperbolic shear polaritons in monoclinic 
gallium oxide via isotopic substitution}
\author[1,*]{Giulia Carini}
\author[2,*]{Mohit Pradhan}
\author[1,*]{Elena Gel\v{z}inyt\.{e}}
\author[3,*]{Andrea Ardenghi}
\author[4]{Saurabh Dixit}
\author[5,6]{Maximilian Obst}
\author[4]{Aditha S. Senarath}
\author[1]{Niclas S. Mueller}
\author[1,7]{Gonzalo Alvarez-Perez}
\author[4]{Katja Diaz-Granados}
\author[4]{Ryan A. Kowalski}
\author[1]{Richarda Niemann}
\author[5]{Felix G. Kaps}
\author[5]{Jakob Wetzel}
\author[2]{Raghunandan Balasubramanyam Iyer}
\author[8]{Piero Mazzolini}
\author[9,10]{Mathias Schubert}
\author[11]{J. Michael Klopf}
\author[12]{Johannes T. Margraf}
\author[3]{Oliver Bierwagen}
\author[1]{Martin Wolf}
\author[1]{Karsten Reuter}
\author[5,6]{Lukas M. Eng}
\author[5,6]{Susanne Kehr}
\author[4]{Joshua D. Caldwell}
\author[1,$\dag$]{Christian Carbogno}
\author[2,$\dag$]{Thomas G. Folland}
\author[3,$\dag$]{Markus R. Wagner}
\author[1,$\dag$]{Alexander Paarmann}

\affil[1]{Fritz Haber Institute of the Max Planck Society, Berlin, Germany}
\affil[2]{University of Iowa, Iowa City, IA, USA}
\affil[3]{Paul-Drude-Institut für Festkörperelektronik, Leibniz-Institut im Forschungsverbund Berlin e.V., Berlin, Germany}
\affil[4]{Vanderbilt University, Nashville, TN, USA}
\affil[5]{TUD Dresden University of Technology, Dresden, Germany}
\affil[6]{Würzburg-Dresden Cluster of Excellence -- EXC 2147 (ct.qmat), Dresden, Germany}
\affil[7]{Center for Biomolecular Nanotechnologies, Italian Institute of Technology, Lecce, Italy} 
\affil[8]{Department of Mathematical, Physical and Computer Sciences, University of Parma, Italy}
\affil[9]{University of Nebraska, Lincoln, NE, USA}
\affil[10]{Lund University, Lund, Sweden}
\affil[11]{Helmholtz-Zentrum Dresden-Rossendorf, Dresden, Germany}
\affil[12]{University of Bayreuth, Bayreuth, Germany}
\affil[*]{These authors contributed equally to this work.}
\affil[$\dag$]{Corresponding authors: carbogno@fhi-berlin.mpg.de, thomas-folland@uiowa.edu, wagner@pdi-berlin.de, alexander.paarmann@fhi-berlin.mpg.de}

\begin{document}

\flushbottom
\maketitle

\begin{abstract}
Hyperbolic phonon polaritons - hybridized modes arising from the ultrastrong coupling of infrared light to strongly anisotropic lattice vibrations in uniaxial or biaxial polar crystals - enable to confine light to the nanoscale with low losses and high directionality. In even lower symmetry materials, such as monoclinic 
\textbeta-Ga$_2$O$_3$ (bGO), 
hyperbolic shear polaritons (HShPs) further enhance the directionality. Yet, HShPs are intrinsically supported only within narrow frequency ranges defined by the phonon frequencies of the host material. Here, we report spectral tuning of HShPs in bGO by isotopic substitution. Employing near-field optical microscopy to image HShPs in \textsuperscript{18}O bGO films homo-epitaxially grown on a \textsuperscript{16}O bGO substrate, we demonstrate a spectral redshift of  $\sim$\,40\,cm\textsuperscript{-1}
for the \textsuperscript{18}O bGO, compared to \textsuperscript{16}O bGO. The technique allows for direct observation and a model-free estimation of the spectral shift driven by isotopic substitution without the need for knowledge of the dielectric tensor. Complementary far-field measurements and \textit{ab initio} calculations - in good agreement with the near-field data - confirm the effectiveness of this estimation. 
This multifaceted study demonstrates a significant isotopic substitution induced spectral tuning of HShPs into a previously inaccessible frequency range, creating new avenues for technological applications of such highly directional polaritons.
\end{abstract}

\begin{multicols}{2}
    
\section{Introduction} 
\label{intro}

Phonon polaritons, quasiparticles resulting from the hybridization of photons with IR-active phonon modes, have recently become a milestone in the field of nanophotonics\cite{caldwell2015low}.
The reduced optical losses, due to the longer phonon lifetimes compared to their plasmonic counterparts, have rendered them a promising toolbox for applications in nanoscale waveguiding\cite{dolado2020nanoscale, he2023guided}, thermal emission\cite{sarkar2024lithography, lu2021engineering}, heat conduction\cite{chen2019modern, pei2023low, pan2023remarkable}, infrared sensing\cite{bylinkin2021real, folland2020vibrational}, and subdiffractional imaging\cite{li2015hyperbolic, dai2015subdiffractional}. In particular, the discovery of hyperbolicity\cite{de2025roadmap}, extensively investigated in materials with hexagonal\cite{caldwell2014sub, li2015hyperbolic, dai2014tunable, li2017optical} and orthorhombic\cite{ma2018plane, alvarez2020infrared, de2021nanoscale, alvarez2022negative} crystal structures, has paved the way for new possibilities to achieve deeply sub-diffractional confinement of light thanks to the large \textbf{k}-vectors made available by the hyperbolic isofrequency contours. In addition, the optical anisotropy arising from the asymmetry in the crystal structure has been shown to be an important factor in enhancing the directionality of polariton propagation\cite{ma2021ghost, ni2023observation}. The existence of a plethora of natural highly anisotropic materials, with different intrinsic properties, has provided a diverse scenario for exploring directional phonon polaritons. In this context, the efforts of the scientific community have recently led to the observation of hyperbolic shear polaritons (HShPs) at the interface of monoclinic crystals\cite{passler2022hyperbolic, hu2023real,matson2023controlling, nunez2025visualization}. HShPs show asymmetric intensity distributions between the two arms of the hyperbola, and dispersive optical axes that are not aligned with the conventional unit cell basis vectors. 
Furthermore, a number of methods for modifying the directionality of HShPs have been demonstrated, e.g., by employing specifically designed nano-antennas\cite{matson2023controlling}, or by placing a twisted thin slab of an orthorhombic van-der-Waals (vdWs) material onto the monoclinic substrate\cite{alvarez2024unidirectional}, inspired by previous studies on twisted \textalpha-MoO\textsubscript{3} bilayers\cite{hu2020topological,chen2020configurable, zheng2020phonon, duan2020twisted}. Additionally, it has recently been shown that shear polaritons can be engineered by stacking slabs of \textalpha-MoO\textsubscript{3} with different thicknesses \cite{zhou2025engineering}. 

Despite all these opportunities, a general hurdle of hyperbolic phonon polaritons is represented by the fact that they are supported only in restricted spectral regions referred to as \textit{reststrahlen bands} (RBs) where the dielectric permittivity has opposite signs along the orthogonal crystal axes (or along the major polarizability axes in the case of monoclinic and triclinic crystals\cite{passler2022hyperbolic}). The typical RBs' widths can range from tens to few hundreds of wavenumbers, depending on the oscillator strengths of the phonon modes involved.
In the broader context of surface phonon polaritons, which share the same limitation of being defined between the material-dependent transverse optical (TO) and longitudinal optical (LO) phonon modes, various ways of overcoming the fixed narrow-band constraint have been explored. To name a few, atomic superlattices consisting of a combination of different polar semiconductors (e.g., GaN, AlN and SiC) with partially overlapping RBs have been realized\cite{ratchford2019controlling,matson2024role}, extending the available spectral range for SPhPs to over 400\,cm\textsuperscript{-1}, which represents approximately 160$\%$ of the RB width of AlN. 
Another example consists in the experimental observation of hybridization between HPhPs and surface plasmon polaritons in vdW heterostructures\cite{dai2015graphene, woessner2015highly, alvarez2022active, zhang2023polariton, garcia2025modulation} 
, which has caused the dispersion curve to change to the point of exceeding 
the upper RB limit. Further modulations of the RB have been induced by ion intercalation\cite{taboada2020broad,wu2020chemical}. 
Finally, a mass-dependent shift in the spectral region of interest can be achieved by a more drastic modification of the material and its properties, such as isotopic substitution\cite{cardona2005isotope}, as so far demonstrated for vdW materials.

Most of the reported work on hyperbolic phonon polaritons in vdW crystals employed thin flakes exfoliated from crystals grown with naturally abundant isotope sources, which typically do not exhibit high degrees of isotopic purity. 
On the other hand, more recent studies investigated the effects of isotopic enrichment in the same vdW materials. In hexagonal boron nitride (hBN), significantly longer phonon lifetimes and, consequently, longer polariton propagation have been reported for isotopically pure crystals\cite{giles2018ultralow}. At the same time, the choice of the boron isotope (\textsuperscript{10}B or \textsuperscript{11}B) enabled a shift of the in-plane TO and LO phonons
\cite{giles2018ultralow} and, thus, of the polariton band by 30\,cm\textsuperscript{-1}.
Building on this prior work, other studies explored the effect of incorporating the \textsuperscript{15}N isotope in various combinations with high concentrations of boron isotopes on the RB shift, however suggesting minimal changes in the losses\cite{janzen2024boron,he2021phonon}.
This concept was also extended to broaden the overall polaritonic band by employing isotopically pure heterostructures\cite{chen2023van}. For biaxial \textalpha-MoO\textsubscript{3}, the existence of ultralow-loss phonon polaritons in Mo-enriched \textalpha-MoO\textsubscript{3} was experimentally demonstrated, in addition to showing a slight red (blue) shift of polariton bands for the \textsuperscript{100}Mo (\textsuperscript{92}Mo) isotope\cite{schultz2024isotopic, zhao2022ultralow}. 

This study moves away from vdW crystals, focusing instead on the effects of isotopic substitution on the polaritons supported by artificially grown 3D crystals, such as bGO\cite{passler2022hyperbolic,matson2023controlling} or calcite\cite{ma2021ghost,ni2023observation}. The high isotopic purity of commercially grown single crystals  of the wide band gap semiconductor bGO is enabling low-loss propagation of polaritons\cite{passler2022hyperbolic}. In this material, therefore, the substitution of this isotope would allow to shift phonons and polaritons to different frequency ranges while maintaining low losses.
Previous work on bGO\cite{ueda1997anisotropy} has shown a significant redshift of approximately 5\,$\%$ for most high-frequency Raman active phonons\cite{janzen2021isotopic} between using naturally abundant \textsuperscript{16}O and \textsuperscript{18}O. 
While similar frequency shifts would be expected for the infrared (IR)-active phonons and the associated polariton bands, this knowledge is still missing.
Understanding the effects of isotopic substitution for this material would allow for better control over spectral tuning of polariton bands, thus widening the applicability of bGO in nanophotonic devices.

\begin{figure*}[ht!] 
\centering\includegraphics[width=1.0\textwidth]{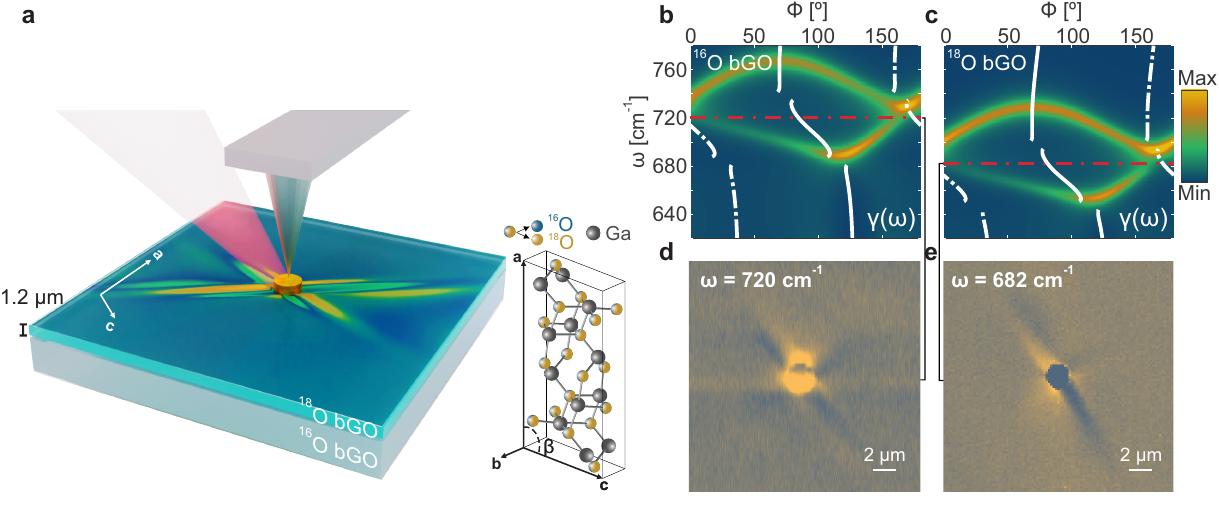}
    \caption{\textbf{Isotopic substitution-induced spectral shift of the hyperbolic shear polaritons in \textbeta-Ga\textsubscript{2}O\textsubscript{3}}. (a) 3D sketch of the Free Electron Laser(FEL)-coupled s-SNOM setup, showing the launching of HShPs by a Au disc on the surface of a 1.2\,\textmu m thick \textsuperscript{18}O bGO film homo-epitaxially grown on a \textsuperscript{16}O bGO substrate. The inset displays a schematic stick-and-ball representation of the conventional unit cell of bGO. The lattice constants were determined to be $a=12.23\,\mbox{\AA}$, $b=3.04\,\mbox{\AA}$, and $c=5.80\,\mbox{\AA}$, while the monoclinic angle defined between the \textit{a} and \textit{c} crystallographic axes is $\beta=103.76^\circ$.\cite{geller1960crystal} The monoclinic \textit{a}-\textit{c} plane coincides with the surface of the sample. (b,c) Azimuthal dispersion for the \textsuperscript{16}O (b) and \textsuperscript{18}O (c) bGO isotopes obtained from transfer matrix simulations at a fixed in-plane momentum $k\textsubscript{ip}/k\textsubscript{0}=1.1$. The white curves show the axial dispersion, that is the frequency dependence of the optical axis direction, $\gamma(\omega)$, which was calculated using eq.\,\ref{eqn:gamma} (see SI Section S6) with the parameters derived from \textit{ab initio} theory (see SI, Table S1). The red dashed-dotted lines indicate the frequencies at which the s-SNOM images shown in panels d,e were acquired. (d,e) Experimental near-field images of HShPs for the \textsuperscript{16}O (d) and \textsuperscript{18}O (e) bGO isotopes taken at incident frequencies of 720\,cm\textsuperscript{-1} and 682\,cm\textsuperscript{-1}, respectively.}
    \label{fig:fig1}
\end{figure*}

In this work, we employ far-infrared near-field optical microscopy as a tool to demonstrate spectral tuning by approximately 40~cm$^{-1}$ of HShPs on the surface of \textsuperscript{18}O bGO thin films homo-epitaxially grown on a \textsuperscript{16}O bGO substrate. 
The employed experimental method enables real-space observation of the propagation of strongly anisotropic HShPs in the frequency range of 660-710\,cm\textsuperscript{-1}. We show that the effect of isotopic substitution can be inferred purely from quantitative analysis of the near-field images, which provides the dispersion of the optical axes as well as the opening angle of the hyperbola for \textsuperscript{18}O bGO. These two quantities, in fact, enable a model-free estimation of the relative spectral shift of the polaritons, as well as of the IR active phonon modes, caused by isotopic substitution, without requiring knowledge of the dielectric tensor of the (less known) \textsuperscript{18}O isotope. The effectiveness of this estimation is verified by comparing with dielectric model fits to infrared reflectance data and first principles calculations.
The ability to evaluate changes in the properties of materials with isotopic purity is particularly important for thin epitaxial films and small scale samples -- where conventional far-field infrared characterization is less accurate\cite{nandanwar2025determining}. Our work introduces and demonstrates isotopic substitution for significant frequency tuning of HShPs in low-symmetry, polar dielectric 3D crystals, paving the way for integrating shear polaritons efficiently within nanophotonic devices. 

\section{Near-field imaging of shear polaritons in isotopically substituted bGO}
\label{exp_res}

The experimental scheme and major results of spectral tuning of HShPs by isotopic substitution are shown in Fig.\,\ref{fig:fig1}. We employed a commercial near-field optical microscope (s-SNOM, \textit{neaSCOPE}) in the self-homodyne detection scheme coupled to a free-electron laser (FEL)\cite{kuschewski2016narrow,helm2023elbe} for near-field imaging of HShPs. As schematically depicted in Fig.\,\ref{fig:fig1}\,a, HShPs are launched off a small Au disc of 2\,\textmu m diameter 
acting as an optical nano-antenna\cite{novotny2009optical,pons2019launching, dai2017efficiency, atkin2012nano} on top of a (0-10) terminated surface of monoclinic \textsuperscript{18}O bGO.
The HShPs are detected interferometrically by measuring the scattered near-field from a metallic tip scanned over an area of 20\,x\,20\,$\mu$m\textsuperscript{2} (see Methods for more details on the experiment). The inset of Fig.\,\ref{fig:fig1}\,a shows a schematic stick-and-ball representation of the conventional unit cell of bGO. The monoclinic $a$-$c$ plane (0-10) is parallel to the surface in our experiment. 
Since $^{18}$O bGO bulk single crystals are not readily available, we employ molecular beam homo-epitaxy\cite{janzen2021isotopic} to grow a strain-free, lattice-matched 1.2~$\mu$m thick $^{18}$O bGO film onto a  commercial \textsuperscript{16}O bGO (0-10) substrate. By employing nano-antennas that only excite  high momenta of the HShPs, the polaritonic near-fields cannot reach the $^{16}$O bGO substrate, and thus, the experiment solely probes the hyperbolic surface modes  of $^{18}$O bGO.

\begin{figure*}[ht!] 
\centering\includegraphics[width=1\textwidth]{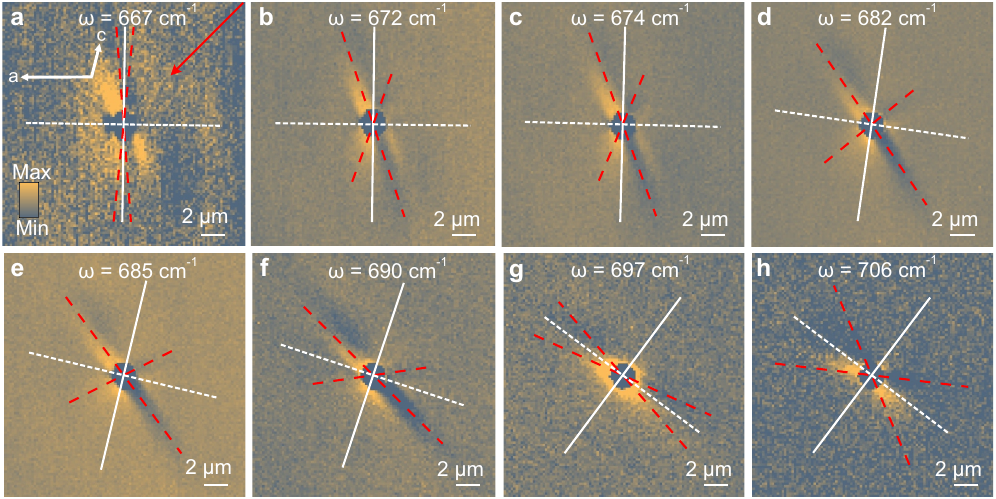}
    \caption{\textbf{Near-field propagation of low-symmetry hyperbolic polaritons in isotopically substituted \textsuperscript{18}O bGO film on \textsuperscript{16}O bGO substrate}. (a-h) Near-field microscopic images of HShPs in a 1.2\,$\mu$m thick isotopically substituted bGO film homoepitaxially grown on a \textsuperscript{16}O bGO substrate. The excitation of the modes occurs though a Au disk with a 2~\textmu m diameter. 
    The panels show images recorded at different excitation frequencies. For all images, we show the optical amplitude demodulated at the second harmonic of the tip tapping frequency (O2A). The signal was acquired in the self-homodyne detection scheme. 
    The red dashed lines indicate the asymptotes of the polariton wave fronts, see SI Section S1 for details on their derivation.   
    The white lines represent the two major polarizability axes of the material system. 
    Panel a contains information on the illumination direction (red arrow) and on the crystallographic axes orientation (white arrows), see SI Section S8.}
    \label{fig:fig2}
\end{figure*}

In the following, we illustrate the spectral tuning of HShPs driven by isotopic substitution, as anticipated from the observed 5\% shift in the Raman active modes\cite{janzen2021isotopic}. To this end, we plot the simulated azimuthal dispersion maps\cite{passler2022hyperbolic,passler2017generalized} for HShPs in Figs.\,\ref{fig:fig1}\,b,c for \textsuperscript{16}O (b) and \textsuperscript{18}O (c) bGO for a fixed in-plane polariton momentum $k\textsubscript{ip}/k\textsubscript{0}=1.1$ 
(thus assuming low confinement), see Methods for details. 
Note that these maps do not consider the epitaxial heterostructure, but were calculated for semi-infinite crystals. Both polariton dispersion plots are notably asymmetric due to the shear effect\cite{passler2022hyperbolic}, arising from an imbalanced intensity distribution across the two arms of the hyperbolic isofrequency contours in momentum space. A close examination reveals that the two plots are nearly identical with the exception of a significant frequency shift of approximately 40\,cm$^{-1}$ between both maps. The spectral shift is further highlighted by the analytical plots of the axial dispersion angle $\gamma(\omega)$, that represents the frequency dependence of the optical axis orientation  (calculated using eq.\,\ref{eqn:gamma}, see SI Section S6), indicated by white curves overlaid with the simulations. To directly probe the resonances (defined in the regions of high intensity in the simulated dispersion maps) would require a prism-coupling experiment\cite{passler2022hyperbolic}. However, due to its intrinsic limitation, this experimental technique can only access low momenta, hence the low confinement assumed in the simulations. 
In the presence of a few \textmu m thick epitaxial layer, as in our experiment, the evanescent fields of surface-bound HShPs would penetrate into the \textsuperscript{16}O substrate, dramatically complicating any analysis. Instead, we employed near-field imaging of the strongly anisotropic HShP propagation launched by nano-antennas, as illustrated in Fig.~\ref{fig:fig1}d,e. This technique allows for access to higher momenta.  Panels d,e show nearly identical near-field patterns 
for both isotopes, yet with a frequency shift of 38\,cm$^{-1}$ between them in agreement with the dispersion plots in Fig.~\ref{fig:fig1} b,c.

In order to quantitatively analyze the spectral tuning of HShPs due to the isotopic substitution, we acquired near-field images of HShPs bound to the air/$^{18}$O bGO interface at several different frequencies as summarized in Fig.\,\ref{fig:fig2}. For these measurements, we keep the azimuthal orientation of the sample fixed with regard to the incident beam as illustrated in Fig.~\ref{fig:fig2}\,a, resulting in a propagation pattern with very large asymmetry arising from intrinsic shear-induced, as well as illumination-induced symmetry breaking\cite{matson2023controlling,hu2023source}. 
Furthermore, the 1.2~\textmu m 
thick \textsuperscript{18}O bGO epilayer 
acts \textit{de facto} as a semi-infinite bulk crystal by supporting surface modes similar to previous work using \textsuperscript{16}O bGO substrates\cite{matson2023controlling}. This is achieved here by employing 2~\textmu m 
diameter Au disks as nano-antennas that efficiently excite HShPs with high momenta ($\sim$\,10\,$k_{\text{0}}$), resulting in evanescent waves with a skin depth of about 0.5\,$\mu$m into the film, see SI Section S9 for details. In contrast, larger antennas excite lower momentum near-fields\cite{matson2023controlling} that penetrate deep enough into the film to reach the substrate, resulting in peculiar patterns that are more difficult to interpret, as discussed in detail in SI Section S3. Furthermore, exciting only large-momentum components of the HShPs also results in a ray-like propagation pattern with the direction of the rays coinciding with asymptotes of the in-plane hyperbolic wave fronts\cite{matson2023controlling,hu2023source}. These experimental design choices significantly reduce the complexity of the s-SNOM images, and enable a thorough quantitative analysis of the real-space polariton propagation -- a non-trivial achievement in the context of low-symmetry polaritons\cite{ma2018plane,ma2021ghost,hu2023source,matson2023controlling}.

The angle between the rays in the real-space polariton propagation is directly linked to the opening angle $\alpha(\omega)$ of the respective in-plane hyperbolic isofrequency surface in momentum space\cite{ma2018plane,alvarez2019analytical}. This is in turn determined by the in-plane anisotropy in the permittivity\cite{alvarez2019analytical} (see eq.\,\ref{alphaan} later in the manuscript). We extracted the direction of both polariton rays from each image, as displayed by the dashed red lines in Figs.\,\ref{fig:fig2}\,a-h, see SI Sections S1.1-S1.3 for details on the fitting procedure. With both rays extracted, we can also experimentally determine the direction of the optical axes, located exactly in the middle between both rays, see solid and dashed white lines in Figs.\,\ref{fig:fig2}\,a-h. Since the rays correspond to the asymptotes of the in-plane hyperbolic wave fronts in momentum space, we can infer that the optical axes coincide with the hyperbola symmetry axes\cite{passler2022hyperbolic}. 
Notably, both parameters -- optical axes and polariton rays directions -- are determined from the near-field data without assuming any physical model, 
and allow for a quantification of the frequency shift from isotopic substitution. Furthermore, these quantities can be linked directly to the in-plane permittivity of $^{18}$O bGO. However, so far the permittivity tensor of the $^{18}$O bGO isotope has not been reported in the literature.

\section{Anisotropic in-plane permittivity of $^{18}$O bGO}
\label{eps_tens}

\begin{table*}[hbt!]
\fontsize{7pt}{7pt}\selectfont
\caption{\textbf{Oxygen isotope effect on in-plane transverse optic phonons in  bGO.} The relative frequency shifts of TO phonons polarized in the monoclinic plane ($\Delta\omega_{rel.}$) were obtained from the experimentally and theoretically derived TO phonon frequencies, as $\frac{\omega_{\text{TO,16}}-\omega_{\text{TO,18}}}{\omega_{\text{TO,16}}}$. A full set of parameters is provided in the SI, Table S1.
}
\label{tab:epsparms}
\centering
\begin{tabular}{l l l l l l l l l l}
\toprule
 & \tabhead{B\textsubscript{u} mode} &  \tabhead{\#1} & \tabhead{\#2} & \tabhead{\#3} & \tabhead{\#4} & \tabhead{\#5} & \tabhead{\#6} & \tabhead{\#7} & \tabhead{\#8} \\
\hline\hline
FT-IR  &  $\Delta\omega_{rel.}$ [\%] & 5.2 & 5.6 & 5.1 & 4.4 & 0.9 & 0.8 & 2.1 & 9.2\\ 
\hline
DFT  &  $\Delta\omega_{rel.}$ [\%] & 5.5 & 5.4 & 4.9 & 4.9 & 1.3 & 1.1 & 3.4 & 4.9\\ 
\bottomrule\\
\end{tabular}
\end{table*}

\begin{figure*}[ht!] 
\centering\includegraphics[width=1\textwidth]{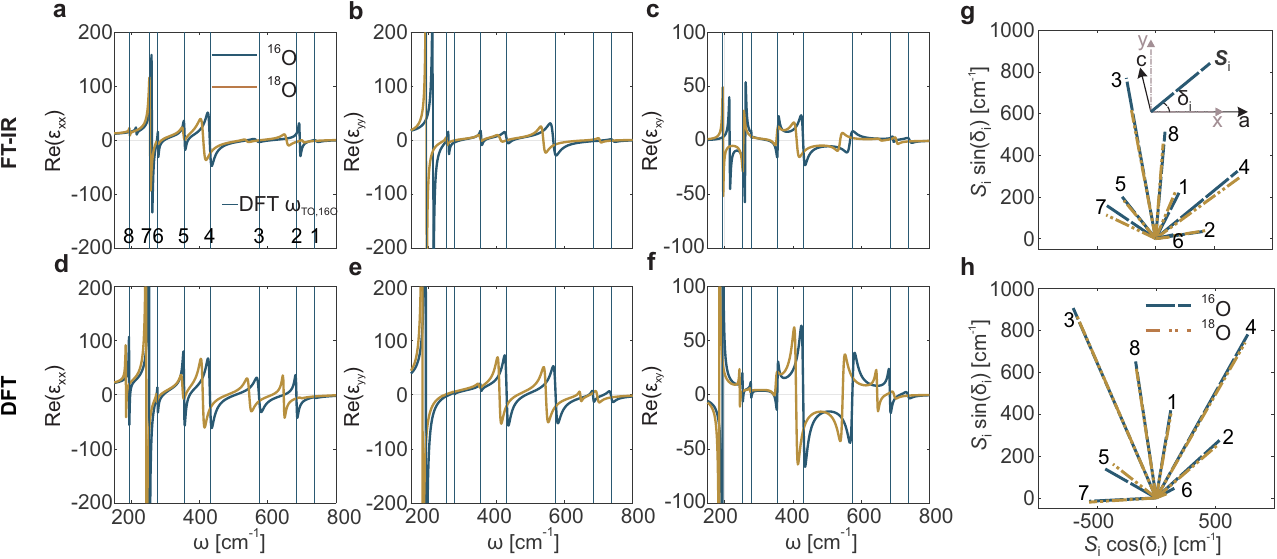}
    \caption{\textbf{Experimental and theoretical permittivity data for the \textsuperscript{16}O and \textsuperscript{18}O bGO}. (a-f) Real parts of $\varepsilon$\textsubscript{xx}, $\varepsilon$\textsubscript{yy}, and $\varepsilon$\textsubscript{xy} for \textsuperscript{16}O and \textsuperscript{18}O bGO, blue and orange lines, respectively, as extracted from azimuth-dependent polarized reflection measurements (a-c) and \emph{ab initio} calculations (d-f). The solid vertical lines indicate the positions of the theoretically derived TO phonon frequencies for the \textsuperscript{16}O bGO isotope. The numbers refer to the IR active modes with \textit{B}\textsubscript{u} symmetry and are consistent with the notation used in Table \ref{tab:epsparms}. (g) Experimentally (FT-IR) and (h) theoretically (DFT) derived oscillator orientation vectors $\boldsymbol{S}_{\text{i}}$ of the \textit{B}\textsubscript{u} modes in the monoclinic $a$-$c$ plane.}
    \label{fig:fig3}
\end{figure*}

To corroborate the near-field extraction of the frequency shift induced by the isotopic substitution, we performed polarized far-field reflectance measurements as well as density-functional theory (DFT) calculations. By means of these two methods, we could experimentally and theoretically determine the in-plane permittivity of $^{18}$O bGO, and compare these results to respective data for $^{16}$O bGO, see Fig.~\ref{fig:fig3}.

The polarized reflectance measurement was conducted with a commercial Fourier-transform infrared (FT-IR) spectrometer (\textit{Bruker Vertex 80V}) to obtain reflectance spectra at multiple azimuthal angles (see Methods for more details\cite{nandanwar2025determining}). The measurements were performed using  the same sample employed for the near-field microscopy [1.2~\textmu m 
\textsuperscript{18}O bGO epitaxially grown on a \textsuperscript{16}O (0-10) bGO substrate], as well as a comparable \textsuperscript{16}O (0-10) bGO substrate. From fits of the reflectivity spectra, we obtain the in-plane permittivity components of \textsuperscript{16}O and \textsuperscript{18}O bGO, shown as orange and blue curves in the plots in Figs.\,\ref{fig:fig3}\,a-c, respectively. 

It is important to note that, because of the monoclinic crystal structure of bGO, the atomic displacements of the $B_u$ phonon modes are not aligned with the crystal axes.  As a consequence, the oscillator orientation vector $\boldsymbol{S}_{\text{i}}$ associated with a given phonon mode forms a non-trivial (neither parallel nor perpendicular) angle $\delta_{\text{i}}$ with the crystallographic axis $a$, as shown in the inset of Fig.\,\ref{fig:fig3}\,g. Note that, within the reference system under consideration, the \textit{a}-axis is oriented in a parallel alignment with the \textit{x}-axis, while the \textit{c}-axis forms an angle of 13.76$^\circ$ 
with the \textit{y}-axis.

The experimentally derived oscillator orientation vectors are plotted in Fig.\,\ref{fig:fig3}\,g; their magnitude is quantified by the oscillator strength, $S_{\text{i}}$. The TO phonon frequencies $\omega\textsubscript{TO,i}$, the oscillator strengths $S_{\text{i}}$, the angles $\delta_{\text{i}}$, and the damping constants $\gamma_{\text{i}}$ were derived from global fit using the commercial software \textit{WVASE} for all eight in-plane IR active modes with \textit{B}\textsubscript{u} symmetry. The full set of fitting parameters along with the corresponding fitting errors is reported in the SI, Table S1, for the \textsuperscript{18}O film and the \textsuperscript{16}O substrate. For more information on the experimental fitting procedure, see SI section S5.

The theoretical results, obtained from the harmonic, anharmonic, and dielectric properties computed via DFT as explained in the Methods, are plotted alongside the experimental data in Figs.\,\ref{fig:fig3}\,d-h. The calculated $\boldsymbol{S}_{\text{i}}$ of the \textit{B}\textsubscript{u} modes for both isotopes are plotted in Fig.\,\ref{fig:fig3}\,h. The full DFT data are reported in SI, Table S1, alongside the experimental values for a direct comparison. 

Overall, the theoretically predicted TO phonon frequencies demonstrate a high degree of correlation with the FT-IR measurements, with maximum deviations of less than
10\,cm\textsuperscript{-1}, comparable to previous work on $^{16}$O bGO\cite{schubert2016anisotropy} (see vertical lines in Fig.\,\ref{fig:fig3}\,a-f). 
Conversely, DFT calculations generally predict larger oscillator strengths in comparison to the experimental parameters. 
In addition, the mode orientations, $\delta_{\text{i,th}}$, predicted by \textit{ab initio} calculations appear rotated around 20$^\circ$ counterclockwise relative to those determined from the FT‑IR reflectance maps, $\delta_{\text{i,exp}}$, as illustrated by the $\boldsymbol{S}$\textsubscript{i}-vector orientation in Figs.\,\ref{fig:fig3}\,g,h. Despite these discrepancies, theory and experiment reveal a comparable trend insofar that the major effect of the isotopic substitution is observed in the phonon frequencies, while the oscillator orientation vectors are largely unaffected. This implies that the HShPs, which arise from the permittivity of the material, will also predominantly experience a spectral shift while propagation characteristics are maintained during isotopic substitution. To quantify the spectral shift of the phonon frequencies, we evaluated the relative spectral shift for the \textit{B}\textsubscript{u} modes, $\Delta\omega_{rel.}$, 
as provided in Table \ref{tab:epsparms}. Consistent between experiment and theory, as well as compared to Raman data\cite{janzen2021isotopic}, we find a spectral shift of approximately 5\,$\%$ 
for the high frequency modes $(\#1-\#4)$, and a smaller shift of only 1\,$\%$ 
in modes $\#$5 and $\#$6, with somewhat poor agreement for the low-frequency modes $\#$7 and $\#$8. This discrepancy can be attributed to the cutoff in the FT-IR reflectance measurements at 200\,cm\textsuperscript{-1}, which affects the fitting results for the damping constants and oscillator strength values for these modes (see SI, Section S5). Nevertheless, the good overall agreement in relative spectral shifts between the far-field and DFT results further validates the impact of isotopic substitution on the dielectric tensor of monoclinic crystals. 

Importantly, we expect the highest frequency modes $\#$1 and $\#$2 to be dictating the permittivity responsible for formation of HShPs observed in Fig.\,\ref{fig:fig2}, suggesting that we can predict a spectral tuning of these modes of approximately 5.5\,$\%$, corresponding to $\Delta\omega\sim$\,40\,cm$^{-1}$ at $\omega=685$\,cm\textsuperscript{-1} in the middle of the hyperbolic polariton band.

\section{Spectral tuning of shear polaritons}

\begin{figure*}[ht!] 
\centering\includegraphics[width=1\textwidth]{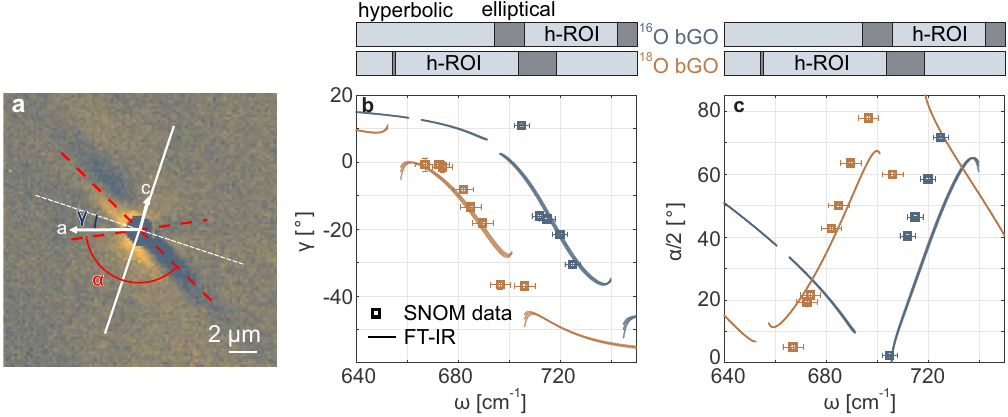}
\caption{\textbf{Experimental verification of the spectral tuning of the optical axis dispersion $\boldsymbol{\gamma}$ and the hyperbola opening angle $\boldsymbol{\alpha}$ by isotopic substitution}. (a) Illustrative sketch of the angles $\gamma$ and $\alpha$. While $\gamma$ is defined between the crystallographic \textit{a}-axis and the major polarizability axis (white dashed line), the opening angle $\alpha$ is defined between the two polariton rays (red dashed lines). As an illustrative example, we have chosen a s-SNOM image for an incident frequency of 690\,cm\textsuperscript{-1}. (b-c) Optical axis dispersion angle $\gamma$ (b) and opening angle $\alpha$ (c) for the \textsuperscript{16}O (blue curve) and \textsuperscript{18}O (orange curve) isotopes of bGO. Squares indicate the near-field results reported with the corresponding error bars (see SI Section S7), extracted from the s-SNOM images in Fig.\,\ref{fig:fig2}, see SI Sections S1-S3 for a full analysis. 
The solid curves have been derived analytically\cite{passler2022hyperbolic, alvarez2019analytical} from the experimental permittivity data (shown in Fig.\,\ref{fig:fig3}), and are reported with their broadening obtained from the experimental uncertainty on the polarized FT-IR measurements (see SI Section S7). The coloured bars above panels b and c show the spectral ranges where different types of phonon polaritons are supported, see SI Fig.\,S7.2. h-ROI indicates the hyperbolic range of interest, where HShPs are defined.} 
    \label{fig:fig4}
\end{figure*}

To further quantify the spectral tuning of HShPs in bGO via isotopic substitution, we compare the angles $\gamma$ and $\alpha$ extracted from the near-field images in Fig.\,\ref{fig:fig2} with those calculated analytically using the FT-IR-extracted permittivity (Fig.\,\ref{fig:fig3}). The results of this comparison are presented in Fig.\,\ref{fig:fig4}. 

While $\gamma(\omega)$ is defined between the crystallographic \textit{a}-axis and the major polarizability axis (white dashed line), the opening angle $\alpha(\omega)$ spans between the two polariton rays (red dashed lines). Fig.~\ref{fig:fig4}\,a contains an illustrative sketch of the two angles.  
In Fig.~\ref{fig:fig4}\,b, we show values for $\gamma$ for $^{18}$O bGO extracted from the data shown in Fig.~\ref{fig:fig2}, as well as for $^{16}$O bGO (see SI Section S4 for the respective near-field images), as orange and blue square symbols, respectively. Indeed, these near-field results show excellent agreement with the corresponding curves for $\gamma$ extracted analytically from the FT-IR obtained permittivity data (Fig.~\ref{fig:fig3}\,a-c) displayed as lines with their broadening indicating the error margins. In contrast to previous studies\cite{passler2022hyperbolic,matson2023controlling}, here the analytical curves for $\gamma$ are calculated using the arctan2 function, which allows for 2$\pi$ periodicity of the optical axis rotation across the entire spectral range: 
\begin{equation}
\label{eqn:gamma}
    \gamma(\omega) = \text{arctan2}(m_{\text{x}},m_{\text{y}})
\end{equation}
The arguments $m_{\text{x}}$ and $m_{\text{y}}$ are the $x$- and $y$-coordinates of the major polarizability axis, $m$, i.e., the principal eigenvector of $\Re(\varepsilon_{\text{xy}})$.

Similarly, we plot the near-field data (square symbols) and the analytical values obtained using the FT-IR permittivity (continuous curves), 
for the half-opening angle $\alpha/2$ in Fig.~\ref{fig:fig4}\,c. To compute the angle $\alpha$ we make use of a generalization of the expression derived for in-plane hyperbolic, biaxial crystals\cite{alvarez2019analytical}:
\begin{equation}
\label{alphaan}
    \alpha(\omega) = 2\,\text{arctan}\left(\sqrt{-\frac{\mathcal{R}{(\varepsilon_{\text{nn}}(\omega))}}{\mathcal{R}{(\varepsilon_{\text{mm}}(\omega))}}}\right)
\end{equation}
where $\varepsilon_{\text{mm}}(\omega)$ and $\varepsilon_{\text{nn}}(\omega)$ are the in-plane permittivity tensor components in the frequency-dependent coordinate system defined by the major polarizability axes\cite{passler2022hyperbolic}. Also for $\alpha$, the near-field observation shows excellent agreement with the analytical prediction (Fig.\,\ref{fig:fig4}\,c). See SI Section S6 for details on the analytical derivation of $\gamma$ and $\alpha$. 

As predicted by \textsuperscript{16}O and \textsuperscript{18}O bGO permittivities from FT-IR measurements, we observe HShPs in the spectral range spannig approximately 40\,cm\textsuperscript{-1}, bonunded by elliptical polaritonic ranges. The hyperbolic and elliptical bands are visualized in Figs.\,\ref{fig:fig4}\,b-c (top), where they are depicted as light-blue and grey shaded horizontal bars, respectively, thereby delineating the hyperbolic range of interest (h-ROI) where we experimentally observed HShPs. The hyperbolic and elliptical bands were identified based on the rotated (in-plane) $\varepsilon$-tensor components\cite{passler2022hyperbolic}, i.e., $\varepsilon$\textsubscript{mm} and $\varepsilon$\textsubscript{nn}, see SI Fig.\,S7.2 for details. Strikingly, the extracted parameters $\gamma$ and $\alpha$ fully characterize the propagation characteristics of the polaritons and, thus, the near-field data shown in Fig.\,\ref{fig:fig4} unambiguously prove the spectral tuning of HShPs, and consequently of the TO phonon frequencies, by isotopic substitution. This finding is based solely on direct observations of polariton propagation in real space.  Although it is useful for proving the technique's effectiveness in quantifying the correct spectral shift, knowledge of the relevant isotope's dielectric tensor is not strictly necessary.

\section{Conclusion}

In summary, we used near-field optical microscopy to study the real-space propagation of surface-bound HShPs in a 1.2\,$\mu$m thick \textsuperscript{18}O  bGO film homoepitaxially grown on a \textsuperscript{16}O bGO substrate. We compared these near-field data with far-field measurements and \textit{ab initio} calculations, showing excellent agreement. The combined and comprehensive study demonstrates a spectral tuning of approximately 40\,cm$^{-1}$ for shear polaritons in bGO using isotopic substitution, while all other polariton characteristics remain largely unchanged. 
Furthermore, our findings show that extracting information on polariton propagation directly from near-field data allows for precise estimation of the spectral shift induced by isotopic substitution without requiring a full measurement of the dielectric tensor. This type of determination is particularly relevant for the analysis of vdW crystals or thin epitaxial layers, where traditional far-field IR spectroscopy has inherent limitations. 
In general, our work demonstrates an exceptional spectral tunability of highly directional HShPs enabled by isotopic substitution, thereby unveiling novel avenues for the exploration of these phenomena at hitherto inaccessible frequencies, and suggests the possibility of its quantification solely with near-field imaging.


\section{Methods}

\subsection*{Sample preparation}

 The \textsuperscript{18}O bGO 
 sample was grown in a plasma-assisted molecular beam epitaxy (PA-MBE) chamber equipped with an RF-plasma source (SPECS PCS). The deposition was made on a 5x5\,mm\textsuperscript{2} \textbeta-Ga$_2$O$_3$ (0-10) Fe-doped substrate purchased from Novel Crystal Technology. Prior to the growth, the substrate was sequentially solvent cleaned using acetone and isopropyl alcohol (IPA) for 5 minutes with sonication, followed by 30\,min of O$_2$ plasma treatment (300\,W, 1\,standard cubic centimeter per minute (SCCM)) 
 in the growth chamber at a substrate temperature of 800\,\textsuperscript{$\circ$}C. The growth was performed at a substrate temperature of 700\,\textsuperscript{$\circ$}C by simultaneously supplying elemental gallium from a double filament effusion cell, with a corresponding beam equivalent pressure (BEP, measured by a nude ion gauge at the substrate position) of BEP\textsubscript{Ga} = 3.4 x 10\textsuperscript{-7}\,mbar, and 1 standard cubic centimeter per minute (1\,SCCM) 
 of isotopically enriched oxygen (\textsuperscript{18}O$_2$) with a purity of 97.39\,$\%$ at 250\,W of RF plasma power. The layer thickness was measured using profilometry. 
 \\

 The Au antennas were fabricated on the surface of the \textsuperscript{18}O bGO 
 sample by a standard electron beam lithography process. A two-layer PMMA resist stack was prepared by sequentially spin-coating AR-P 662.04 and AR-P 679.04 over an adhesion promoter (AR 300-80) followed by soft bake. A conductive Electra coating was applied to prevent charging during exposure. After exposure, the conductive layer was rinsed off, and the resist was developed with AR 600-56 developer, followed by IPA. Metal deposition of Cr/Au (5\,nm / 50\,nm) was followed by lift-off in acetone in an ultrasonic bath.

\subsection*{FEL s-SNOM measurements}

The near-field images illustrated in Fig.\,\ref{fig:fig2} were acquired by means of a commercial s-SNOM system from \textit{neaSPEC} (now part of \textit{attocube systems}) coupled to the free-electron laser FELBE 
at Helmholtz Zentrum Dresden-Rossendorf (HZDR).
FELBE was operated at a repetition rate of $\sim$13 MHz, with a spectral bandwidth of about 0.5\%. For the complete dataset, the incident light was p-polarized. The measurements were performed using a mercury-cadmium-telluride (MCT) detector in a self-homodyne detection scheme, due to the comparatively low signal-to-noise ratio of the light source that does not allow reliable operation using the pseudo-heterodyne method. In this configuration, the detected signal contains a mixture of near-field and background contributions and the optical amplitude and phase are mixed and cannot be separated\cite{chen2019modern}. Still, an analysis of the features of the HShPs separate to the background contributions was possible due to their high-momentum nature compared to the low momenta dominating the background. 
More details on the particular setup are available in previous publications\cite{matson2023controlling, obst2023terahertz}.

\subsection*{Transfer Matrix}

The azimuthal dispersion maps in Figs.\,\ref{fig:fig1}\,b-c were 
derived from transfer matrix (TM) simulations\cite{passler2017generalized}, a formalism that allows for numerical evaluation of the reflection and transmission coefficients in multilayered media. The quantity plotted is the imaginary part
of the Fresnel reflection coefficient for p-polarized
light, $\Im(R\textsubscript{pp})$, normalized to its maximum. As
shown in a previous work\cite{passler2023layer}, the high intensity regions emerging from these plots are suitable for potential excitation of surface-bound polaritonic modes, since the maxima of $\Im(R\textsubscript{pp})$ coin-
cide with the material’s resonant modes.

\subsection*{FT-IR measurements}
The Fourier Transform Infrared (FT-IR) spectroscopy measurements were performed for the \textsuperscript{16}O bGO substrate and the isotopically enriched \textsuperscript{18}O bGO film sample using a commercial FT-IR spectrometer (Bruker Vertex 80V) with a modified sample holder (A513/QA). Polarized reflectance spectra for both samples were measured at varying azimuthal orientations, ranging from 0$^\circ$ to 180$^\circ$  with 5$^\circ$  increments. The incidence angle was fixed for each azimuthal measurement: 50$^\circ$  for \textsuperscript{16}O bGO, and 15$^\circ$  for the \textsuperscript{18}O bGO epitaxial layer. The optical measurement layout consists of a deuterated L-alanine doped triglycine sulphate (DLaTGS) IR detector, a KRS5 polarizer, and a Si beam splitter.

\subsection*{Dielectric tensor formula for monoclinic crystals}

The in-plane dielectric tensor components for monoclinic crystals are derived from the Lorentz oscillator model through the following equation:\cite{schubert2016anisotropy}:

\begin{equation}
\label{Lorentz_model}
    \boldsymbol\varepsilon (\omega) = \boldsymbol \varepsilon_{\infty} + \sum_i^{N_\textrm{phonon}}\frac{\boldsymbol S_{i} \times \boldsymbol S_{i}^\ast}{\omega_{\textrm{TO},i}^2 - \omega^2 - i\omega \gamma_i}
\end{equation}
where $\varepsilon_{\infty}$ denotes the electronic contribution to the dielectric permittivity, $\boldsymbol S_{i}$ the oscillator orientation vector, $\omega_{\textrm{TO},i}$ the transverse optical phonon frequency, and $\gamma_i$ the damping constant. The summation is performed over the eight IR active phonon modes that are defined in the monoclinic plane. The experimentally derived parameters for the dielectric tensor components plotted in Fig.\,\ref{fig:fig3}, using eq.\,\ref{Lorentz_model}, are reported in SI, Table S1. The angle between the $x$-axis and the vector $\boldsymbol S_{\text{i}}$ associated with the phonon mode $i$ is computed as $\delta_i  = \arctan{(S_{\text{i,y}}/S_{\text{i,x}})}$. It should be noted that the oscillator strengths, $S_{\text{i}}$, are expressed in units of cm\textsuperscript{-1}.

\subsection*{DFT calculations}
To assess the properties of phonon polaritons in bGO from first principles, all parameters entering the Lorentz oscillator model defined by eq.\,(\ref{Lorentz_model}) are calculated via DFT viz.~density-functional perturbation theory~(DFPT)~\cite{gonze1997dynamical} using the FHI-aims code~\cite{blum2009ab,Abbott2025}. All presented results were obtained using a $X \times X \times X$ \textbf{k}-grid in the first Brillouin zone, ``tight'' defaults for the basis set and the numerical settings~\cite{blum2009ab,Carbogno.2022}, and the local-density approximation in the parametrization proposed by Perdew and Zunger~\cite{perdew1981self} for modeling exchange and correlation, in line with earlier DFT studies of bGO~\cite{schubert2016anisotropy}.

More specifically, bGO was first relaxed under symmetry constraints~\cite{Lenz.2019} until the largest  (generalized) force component stemming from atomic and lattice degrees of freedom fell below 0.001 eV/$\mathring{\textrm{A}}$. For this optimized structure, the electronic contribution to the
dielectric permittivity~$\boldsymbol \varepsilon_{\infty}$ was calculated with DFPT~\cite{Shang.2018ucl}, while the harmonic phonon properties (including the angular frequencies $\omega_{\textrm{TO},i}$) and the first-order anharmonic corrections defining the damping constants  $\gamma_i$ were computed using finite displacements in a $4\times2\times2$ supercell. To this end, the FHI-vibes\cite{Knoop.2020fmh} interface to the implementations available in Phonopy\cite{togo2023first} and Phono3py\cite{togo2015distributions} was used. The thereby obtained eigenvectors~$\mathbf{\hat e}_{i \kappa}$  of the dynamical matrix at $\Gamma$ were used to compute the oscillator orientation vectors defined as~\cite{gonze1997dynamical}
\begin{equation}
    \boldsymbol S^{\textrm{DFT}}_{\text{i}} = \frac{1}{c\sqrt{\varepsilon_0 V}} \sum_\kappa^{N_{\textrm{atoms}}}\frac{1}{\sqrt{M_\kappa}}\mathbf{Z}_\kappa \mathbf{\hat e}_{\text{i} \kappa} ,\label{eq:Sdft}
\end{equation}
where $c$ denotes the speed of light,  $\varepsilon_0$ the free-space permittivity, $V$ the volume of the unit cell, $M_\kappa$  the mass of the atom with index $\kappa$, and $\mathbf{Z}_\kappa$ its Born effective charge (tensor). The latter were again computed via finite differences,~i.e.,~by monitoring the change in polarization~\cite{Carbogno.2025} upon the displacement of atom $\kappa$~\cite{Akkoush.2024}. By using these definitions, both $\mathbf{S}_{\text{i}}$ and $\boldsymbol\varepsilon (\omega)$ are directly comparable between theory and experiment; results for isotopic substitution can be straightforwardly obtained by adapting the atomic masses $M_\kappa$ in Eq.~(\ref{eq:Sdft}) and in the (an)harmonic phonon calculations.

\section*{Acknowledgements}
MBE growth was performed in the framework of GraFOx, a Leibniz-ScienceCampus, and was funded by Deutsche Forschungsgemeinschaft (DFG, German Research Foundation)-Project No. 446185170. Nicole Volkmer is gratefully acknowledged for fabrication of the Au disc launchers. EG thanks the Alexander von Humboldt foundation for research fellowship funding. Parts of this research were carried out at the ELBE Center for High-Power Radiation Sources at the Helmholtz-Zentrum Dresden-Rossendorf e.V., a member of the Helmholtz Association. M.O., F.G.K., J.W., L.M.E., and S.C.K. acknowledge the financial support by the Bundesministerium f\"ur Bildung und Forschung (BMBF, Federal Ministry of Education and Research, Germany, Project Grant Nos. 05K19ODA, 05K19ODB and 05K22ODA) and by the Deutsche Forschungsgemeinschaft (DFG, German Research Foundation) under Germany's Excellence Strategy through W\"urzburg-Dresden Cluster of Excellence on Complexity and Topology in Quantum Matter-ct.qmat (EXC 2147, project-id 390858490) and through the Collaborative Research Center on Chemistry of Synthetic Two-Dimensional Materials-CRC1415 (ID: 417590517). 
K.D.-G. was supported by the Department of Energy - Basic Energy Sciences under Grant number DE-FG02-09ER4655, while R.A.K acknowledges that this work was supported by a NASA Space Technology Graduate Research Opportunity. J.D.C. and S.D. was supported by the Office of Naval Research  MURI on Twist-Optics under grant N0001-23-1-2567. N.S.M. acknowledges funding from the Deutsche Forschungsgemeinschaft
(DFG, German Research Foundation) - Projektnummer 551280726.

\bibliographystyle{MSP}
\bibliography{bibliography}

\end{multicols}

\clearpage

\end{document}


\flushbottom
\maketitle

\section{s-SNOM image analysis}

In this section, we provide a detailed description of the data treatment performed to derive the frequency-dependent rotation angle $\gamma$ and the opening angle $\alpha$ from the s-SNOM images shown in the main text (Fig.\,2). The experiment was carried out using the self-homodyne detection scheme, i.e. the reference arm of the interferometer in the s-SNOM setup was blocked during the scans. This detection scheme provides only an intensity measurement, while the near-field phase information cannot be decoupled from the background phase\cite{chen2019modern}. The plots in Fig.\,2 display the amplitude of the optical signal demodulated at the 2\textsuperscript{nd} harmonic of the cantilever oscillation frequency. We found the demodulation at the second harmonic to be sufficient to extract the near-field contribution and all optical images analyzed in this work were demodulated at this harmonic. 
In order to obtain a quantitative estimate of the polariton orientation from these images, several levels of analysis were performed that are exemplified in the following for an incident wavelength of 682\,cm$^{-1}$.

\subsection{Identification of Au discs position}

\begin{figure}[ht!] 
\centering\includegraphics[width=0.7\textwidth]{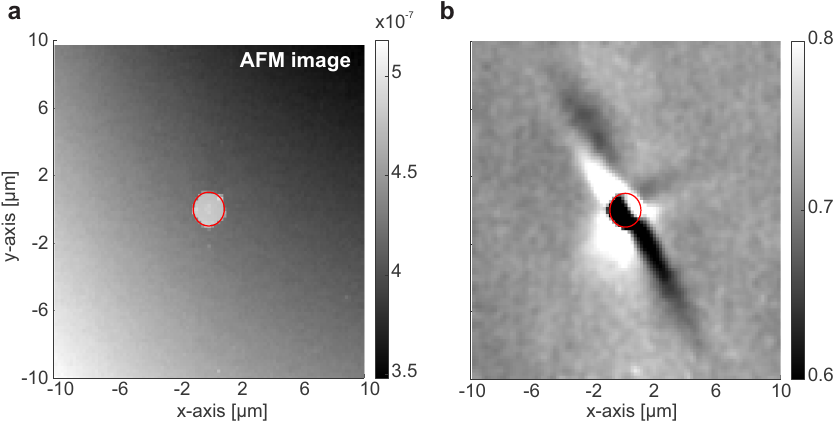}
    \caption{\textbf{Identification of Au disc position}. (a,b) AFM (a) and s-SNOM (b) images after being re-centred. The pixels were shifted by the disc centre coordinates $(x_{\text{0}},y_{\text{0}})$. Both images are overlaid with a circle of radius 1\,\textmu m centred on the origin (red solid line).}
    \label{fig:figS1}
\end{figure}

First, the position of the Au discs 
was determined to identify the region where the polariton is launched. This was achieved by analyzing the atomic force microscopy (AFM) images provided by the \textit{neaSCOPE} device
(from \textit{neaSPEC}, now part of \textit{attocube systems}) 
simultaneously with the near-field scans; this allows fair referencing. To find the coordinates of the centre of the disks $(x_{\text{0}},y_{\text{0}})$, we identified the rising and falling edges of the peaks in the topography using Matlab's 'findpeaks' function and calculated the midpoint between them. All pixels in the AFM and optical images were then shifted accordingly to re-centre the disc. In Figs.\,\ref{fig:figS1}\,a-b we have superimposed the corrected AFM (a) and near-field optical images (b) with a circle (red solid line) of radius 1\,\textmu m centred at the origin.

\subsection{Analysis of the polariton propagation direction}
\label{pol_prop_dir}

\begin{figure}[ht!] 
\centering\includegraphics[width=1\textwidth]{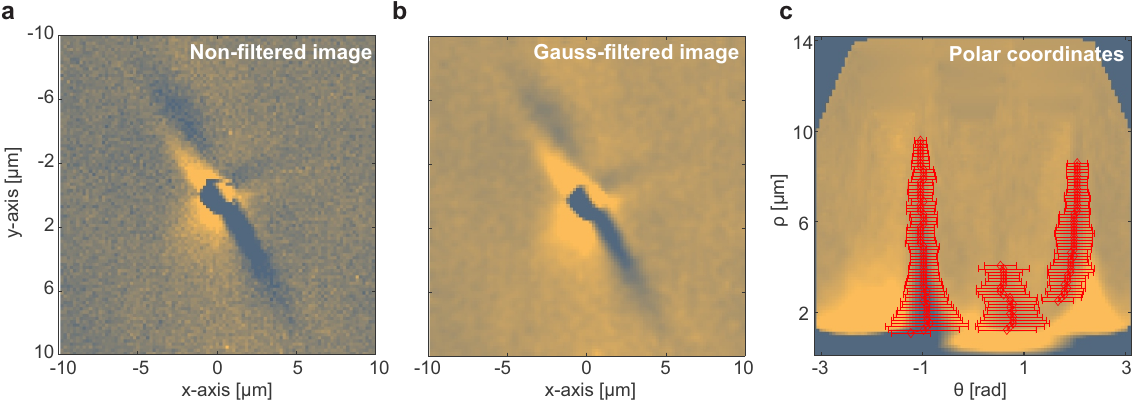}
    \caption{\textbf{Analysis of the polariton propagation direction}. (a-b) Raw (a) and gauss-filtered (b) near-field optical images demodulated at the second harmonic for an incident wavelength of 682\,cm$^{-1}$. The pixels had been shifted by the disc centre coordinates $(x_0,y_0)$ (see Fig.\,\ref{fig:figS1}). (c) Optical image converted to polar coordinates, overlaid with the dip positions of the longest segments for each 'angular' region. The error bars (red solid lines) correspond to the FWHM of the extracted dips.}
    \label{fig:figS2}
\end{figure}

Fig.\,\ref{fig:figS2} describes the subsequent steps of analysis. The raw near-field scan  
for an incident frequency of 682\,cm$^{-1}$ is shown in panel a, while in panel b a gauss filter was applied to the image to smooth the data, reducing the noise and simplifying the following data treatment. Note that the polariton in Figs.\,\ref{fig:figS2}\,a-b is launched from a Au disc of 2\,\textmu m diameter. As shown in a previous paper\cite{matson2023controlling}, the size of the antenna acts as a momentum filter and plays an important role in determining the anisotropy of the resulting polariton wave (see Section \ref{disc_size}). Small discs, like the one we chose for our measurements, allow the selection of higher momenta. Due to the asymmetric intensity distribution along the two arms of the hyperbola in momentum space, this leads to a ray-like polariton propagation in real space\cite{alvarez2024unidirectional}. The absence of fringes simplifies the determination of the polariton rays direction, which we derive from the dips in the near-field optical intensity signal ($S_2$).

Since the polariton waves are emitted from the edges of the antenna, the rays we observe are expected to be orthogonal to a circle centred on the origin. Therefore, the easiest method to extract the polariton propagation direction is to treat the rays as vertical lines in the near-field optical image, once converted from Cartesian to polar coordinates, as shown in Fig.\,\ref{fig:figS2}\,c. The conversion was performed by first transforming the coordinates $(x,y)$ of each pixel of the image in panel b into the corresponding polar coordinates $(\rho,\theta)$ and then interpolating (with Matlab's function 'scatteredInterpolant') for new axes $\rho_{\text{eq}}$ and $\theta_{\text{eq}}$ with equidistant points. 

This coordinate transformation allowed for a more straightforward extraction of the dip positions in the near-field intensity signal. The image was divided into three different angular regions (corresponding to a short and two long rays) and horizontal line cuts were applied at each $\rho_{\text{eq}}$ value. By applying Matlab's 'findpeaks' function to $\text{Max}(S_{\text{2}})-S_{\text{2}}$ (where S\textsubscript{2} is the intensity of the near-field optical signal demodulated at the second harmonic), we were able to extract the near-field intensity dips along with their full width at half maximum (FWHM).
To ensure that the false dips were discarded, we set appropriate thresholds for minimum peak prominence and maximum peak width, and then selected the dip with the largest peak prominence among those extracted by the function. Thus, for each angular region, we obtained an array of corresponding dip positions and FWHMs. If the criteria mentioned above were not met, NaNs were written as array elements instead. This led us to define several segments between the NaNs, among which the longest one was selected for each polariton ray. The dip positions along with the corresponding FWHM are shown in Fig.\,\ref{fig:figS2}\,c, overlaid with the result of the interpolation discussed above. 

\subsection{Analysis extended to various incident FEL frequencies}
\begin{figure}[ht!] 
\centering\includegraphics[width=1\textwidth]{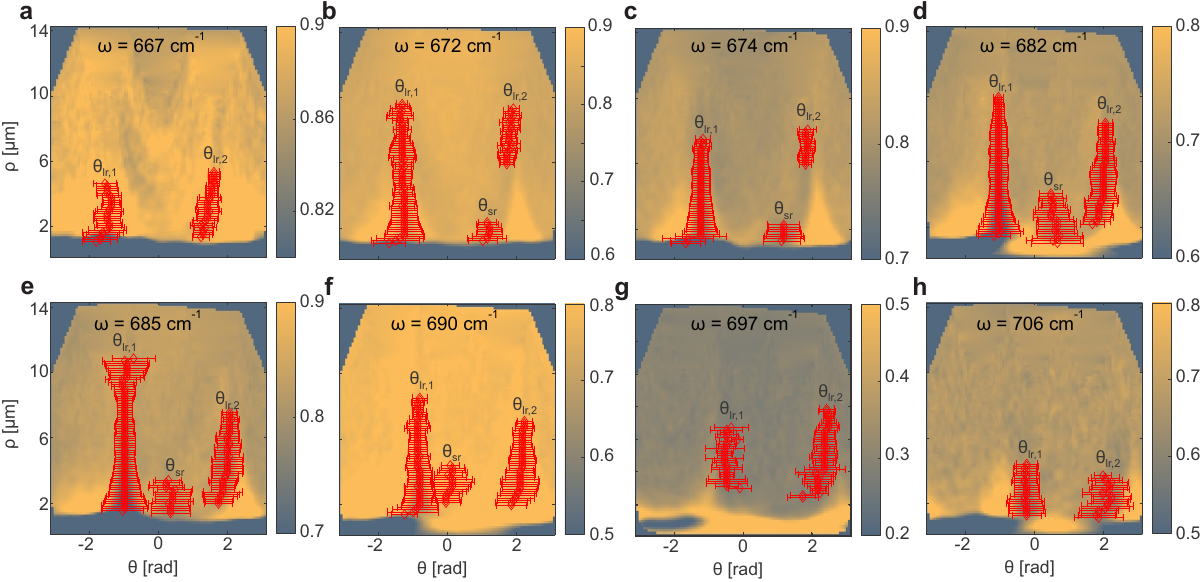}
    \caption{\textbf{Extension of analysis procedure to various frequencies}. (a-h) Near-field optical images converted to polar coordinates, overlaid with the dip positions of the longest segments for each angular region. Each panel corresponds to a different frequency of the FEL incident light. The error bars (red solid lines) correspond to the FWHM of the extracted dips.}
    \label{fig:figS3}
\end{figure}

The procedure illustrated in the previous sections for one frequency of the incident FEL beam is now applied to various frequencies. Fig.\,\ref{fig:figS3} shows the s-SNOM images after conversion into polar coordinates, interpolation for the axes $\rho_{\text{eq}}$ and $\theta_{\text{eq}}$, and derivation of the dip positions.
Although the analysis procedure was automated, for each FEL incident frequency we selected different areas in which to examine the dips and set different thresholds for minimum peak prominence and maximum peak width. In particular, looking at the s-SNOM images in Fig.\,2 of the main text, we cannot detect the presence of a short ray for all FEL incident frequencies: for $\omega_{\text{inc}}\,=\,$[667,\,697,\,706]\,cm$^{-1}$, the short ray cannot be distinguished from the long ray in the upper half of the images. Therefore, in these cases we consider only two angular regions (instead of three). The dip positions, plotted as red circles in Fig.\,\ref{fig:figS3}, were then averaged with a weighted mean using the peak prominence values as weights. This allowed us to determine the angle at which the rays propagate, which we then used to plot the red dashed lines in Fig.\,2 of the main text.

\section{Experimental derivation of optical axis dispersion angle $\gamma$ and opening angle $\alpha$}

This section elucidates the procedure followed to extract the experimental data shown in Fig.\,4 of the main text. Experimentally, the angles $\gamma$\textsubscript{exp} and $\alpha$\textsubscript{exp} were obtained after applying the whole procedure described in the previous sections S1-S3, and identifying two or three main rays in the s-SNOM images. Let us consider Fig.\,\ref{fig:figS3}. In panels b-f we could recognize three main areas where the near-field optical signal has clear minima. We identified two long rays and one short ray, due to the strong optical anisotropy of the underlying crystal. For these incident frequencies, $\alpha$\textsubscript{exp} and $\gamma$\textsubscript{exp} were calculated as follows:

\begin{equation}
\begin{aligned}
\label{alphaexp}
\alpha\textsubscript{exp} = \pi - |\theta\textsubscript{lr,1} - \theta\textsubscript{sr}| 
\end{aligned}
\end{equation}

\begin{equation}
\begin{aligned}
\label{gammaexp}
\gamma\textsubscript{exp} = \theta\textsubscript{lr,1} - \alpha\textsubscript{exp} /2\\
\end{aligned}
\end{equation}
In Fig.\,\ref{fig:figS3}\,a,g,h the short ray is not visible, since  these frequencies are at the limits of the hyperbolic region, where we expect the arms of the hyperbola to converge. This transition can be observed in Fig.\,\ref{fig:figS3}: while still absent in panel a, the short ray starts appearing in panel b from the long ray at $\theta$\textsubscript{lr,2}, from which it becomes increasingly separated at increasing frequencies (i.e., in the next panels), to finally merge again with the other long ray, $\theta$\textsubscript{lr,1} in panel h. To account for this behaviour, we had to make two different cases for the two ends of the hyperbolic band: for panel a, $\alpha\textsubscript{exp} = \pi - |\theta\textsubscript{lr,1} - \theta\textsubscript{lr,2}| $, and $\gamma$\textsubscript{exp} was calculated as in eq.\,\ref{alphaexp}, while for panels g,h $\alpha\textsubscript{exp} = |\theta\textsubscript{lr,1} - \theta\textsubscript{lr,2}| $ and $\gamma\textsubscript{exp} = \theta\textsubscript{lr,1} - ( \pi - \alpha\textsubscript{exp} /2)$.

\section{Comparison with larger disc size} 
\label{disc_size}

In this section we compare the main results with the data obtained using a disc with a larger diameter of 4\,\textmu m. As shown in a previous publication\cite{matson2023controlling} and mentioned above (see section \ref{pol_prop_dir}), the disc size plays an important role in determining the anisotropy in the polariton propagation. The directionality of the mode decreases with disc diameter: when launched from a 4\,\textmu m diameter disc, the polariton propagates with clearly visible tilted wave fronts instead of a ray-like behaviour. These wave fronts are useful in determining the polariton propagation direction, which is shown by the white solid lines in Fig.\,2.

\begin{figure}[ht!] 
\centering\includegraphics[width=0.7\textwidth]{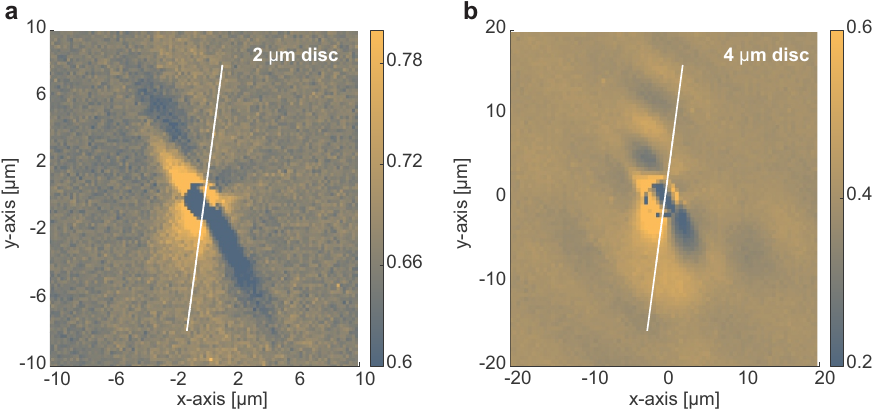}
    \caption{\textbf{Comparison between polaritons emitted by discs with different diameters}. (a-b) Second harmonic of the near-field optical signal for Au discs with diameters of 2 \textmu m (a) and 4 \textmu m (b) at an incident wavelength of 682\,cm$^{-1}$. The solid white line indicates the polariton propagation direction derived from the analysis procedure discussed in the previous sections.}
    \label{fig:figS4}
\end{figure}

In Fig.\,\ref{fig:figS4}, the raw data obtained with the 2\,\textmu m diameter disc (a) is compared with the near-field optical signal produced by a 4\,\textmu m diameter disc (b). While panel a shows a well-defined polariton propagating from the Au disc in a ray-like fashion, the pattern traced by the polariton in panel b appears more complex to interpret. The larger momentum modes launched by the smaller disc decay more rapidly, and the resulting patterns are free of artifacts (as displayed in Fig.\,\ref{fig:figS4}\,a). On the contrary, for the 4\,\textmu m diameter disc, we can distinguish two different fringe patterns: 

\begin{enumerate}
\item The interference pattern with a shorter wavelength emerges at the air/\textsuperscript{18}O bGO film interface due to the lower asymmetry in propagation with respect to polaritons launched by smaller discs, which makes the wave fronts more pronounced, similar to previous observations in \textsuperscript{16}O bGO bulk single crystals\cite{matson2023controlling}.
\item The other interference pattern component is due to the larger penetration depth of the low momentum modes into the film, which allows them to couple to the film/substrate interface.
\end{enumerate}

Despite the complexity of the image, the shorter wavelength fringes (corresponding to the wave fronts of the hyperbolic shear polariton launched at the air/\textsuperscript{18}O bGO film interface) can be useful to determine the direction of propagation. In fact, we have chosen $\alpha$\textsubscript{exp} to be defined as in eq.\,\ref{alphaexp}, but we could just as well have defined this angle as $\pi-\alpha_{\text{exp}}$. In this case, $\gamma$\textsubscript{exp} would be rotated by $\pi/2$. The comparison illustrated in Fig.\,\ref{fig:figS4} shows that the ambiguity can be overcome when the polariton wave fronts are taken into account.

\section{Study of hyperbolic shear polaritons in $^{\text{16}}$O bGO} 
\label{16O}

\begin{figure}[ht!] 
\centering\includegraphics[width=1\textwidth]{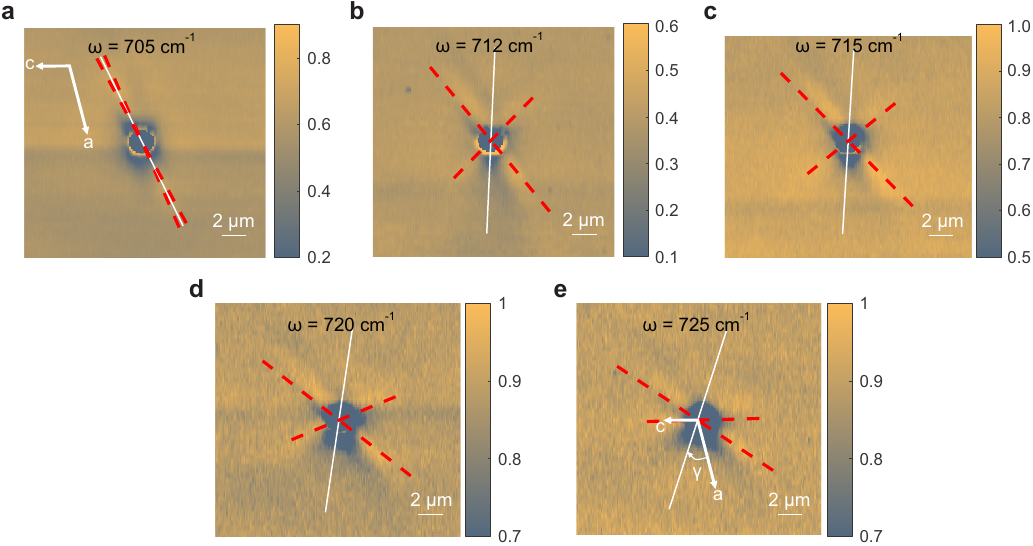}
    \caption{\textbf{Hyperbolic shear polariton real-space propagation in $^{\text{16}}$O bGO}. (a-e) Second harmonic of the near-field optical signal for various wavenumbers in a $^{\text{16}}$O bGO substrate.  The polaritons are launched by Au discs with a diameter of 2 \textmu m. The red dashed lines indicate the orientation of the polariton rays, while the white solid line illustrates the propagation direction. Panel e reports a sketch of the crystal axes and shows how the angle $\gamma$ is defined relative to them.}
    \label{fig:figS5}
\end{figure}

In this section, we study the propagation of hyperbolic shear polaritons (HShPs) on a $^{\text{16}}$O bGO substrate. The experimental data points (blue squares) displayed in Figs.\,4\,b-c of the main text are extracted from the s-SNOM images in Fig.\,\ref{fig:figS5}. To account for artifacts caused by the mixing of amplitude and phase channels resulting from the self-homodyne detection scheme employed in the experiment, we multiply our (intensity) images with the cosine of the phase of the lock-in amplifier\cite{obst2023terahertz,matson2023controlling}, as follows:

\begin{equation}
   S_{\text{2}} = - \text{O2A} \cdot \text{cos(O2P)}.
\end{equation}

The signal is demodulated at the second harmonic. The HShPs shown in Fig.\,\ref{fig:figS5} are launched by a Au disc with a 2 \textmu m diameter. Furthermore, a circular mask with a radius of 1 \textmu m centered on the disc has been applied to each s-SNOM image. The data are analyzed as described in detail in the previous sections of this SI. The red dashed lines are derived from the maxima in the distribution of the near-field signal for the long and the short ray. Note that in Fig.\,\ref{fig:figS5}\,b-e, we have selected only one quadrant for the long ray and one quadrant for the short ray. As we can see in the respective panels, the signal seems to undergo a phase flip in the opposite quadrant, causing the maxima to become minima. Therefore, the red dashed lines derived from the analysis overlap with the brightest regions of the images only in two quadrants. In Fig.\,\ref{fig:figS5}\,a, the short ray is not visible, because the polariton is canalized there due to the presence of a topological transition between a hyperbolic and an elliptical spectral regimes. The white solid line indicates the direction of the polariton propagation. Note that the angle $\gamma$\textsubscript{exp} (as plotted in Fig.\,4 of the main text) is defined with respect to the crystal axes as shown in Fig.\,\ref{fig:figS5}\,e.

\section{Experimental characterization of the dielectric permittivity tensor}
\label{expepsilon}

The experimental parameters for the dielectric tensor components plotted in Fig.\,3\,a-c were derived from polarized reflectance measurements carried out using Fourier transform infrared (FTIR) spectroscopy. The measurements were performed (1) on a reference [010] \textsuperscript{16}O bGO substrate, and (2) on the isotopically substituted \textsuperscript{18}O bGO film homoepitaxially grown on a \textsuperscript{16}O bGO substrate, utilizing a commercial FT-IR spectrometer (Bruker \textit{Vertex 80V}) with a modified sample holder (A513/QA bench modified). The sample’s reflectance was measured at a fixed angle of incidence of 15$^\circ$ for the \textsuperscript{18}O bGO film, and at 50$^\circ$ for the reference \textsuperscript{16}O bGO bulk crystal. The smaller angle of incidence was used for the \textsuperscript{18}O bGO film to minimize the influence of the out-of-plane permittivity, which cannot be sufficiently accurately determined from a single crystal cut. The azimuthal orientation was varied from 0$^\circ$ to 180$^\circ$ at an increment of 5$^\circ$ for both isotopes. The detection scheme employed in the measurements consists of a deuterated L-alanine doped triglycine sulphate (DLaTGS) IR detector, a KRS5 polarizer, and a Si beam splitter. 

\begin{figure}[ht!] 
\centering\includegraphics[width=0.8\textwidth]{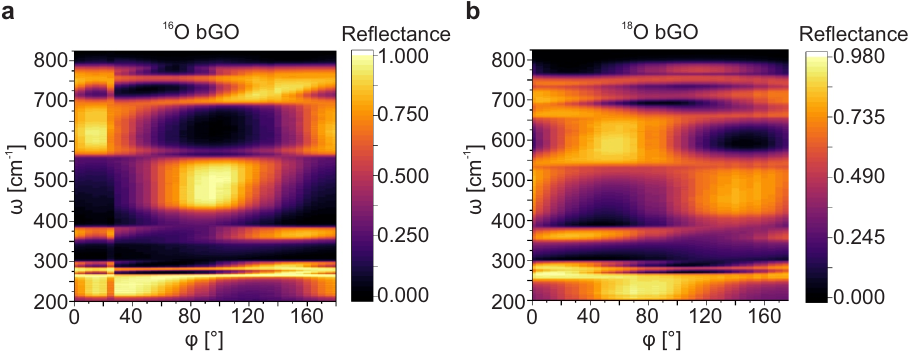}
    \caption{\textbf{FT-IR reflectance contour maps obtained from the raw data}. (a-b) FT-IR reflectance contour maps for the \textsuperscript{16}O bGO bulk single crystal (a) and the \textsuperscript{18}O bGO 1.2\,$\mu$m thick film homoepitaxially grown on the \textsuperscript{16}O bGO substrate (b).}
    \label{fig:figS5_1}
\end{figure}

The reflectance data for the respective materials are represented as reflectance contour maps in Fig.\,\ref{fig:figS5_1}. These results show the polarization dependent/independent RBs \cite{schubert2019phonon} for the two materials: the polarization independent RBs exist for a given frequency at all azimuthal angles, while the polarization dependent ones only exists within a specific azimuthal range. The maps demonstrate how the isotopic substitution has modified the RBs within the IR spectra. However, it is important to note that in the data shown in Fig.\,\ref{fig:figS5_1}\,b both the \textsuperscript{18}O bGO epilayer and the \textsuperscript{16}O bGO substrate play a significant role.

The dielectric permittivity tensor was fitted using a multi-oscillator model, as first described in Ref.\,\cite{schubert2016anisotropy}, using \textit{WVASE} 
\cite{WVASE}. \textit{WVASE} is a proprietary software for analyzing ellipsometry, transmission and reflectance data. The initial fitting was done for the \textsuperscript{16}O bGO reflectance data using literature values from Ref.\,\cite{schubert2016anisotropy} of the IR active modes with B\textsubscript{u} and A\textsubscript{u} symmetry:
the TO phonon frequencies $\omega\textsubscript{TO,i}$, the oscillator strengths $\lVert$\textbf{S}$_{\text{i}}\rVert$, the phonon angles $\delta_{\text{i}}$, and the damping constants $\gamma_{\text{i}}$ taken from the aforementioned reference served as initial parameters in the harmonic Lorentz oscillator model (see Methods in the main text) to fit the reflectance data. Before optimizing the parameters to fit the experimental data, we applied a $\phi$ Euler rotation of 65.5$^\circ$ in the $xy$-plane of the laboratory frame to match the literature data. 


\begin{figure}[ht!] 
\centering\includegraphics[width=0.9\textwidth]{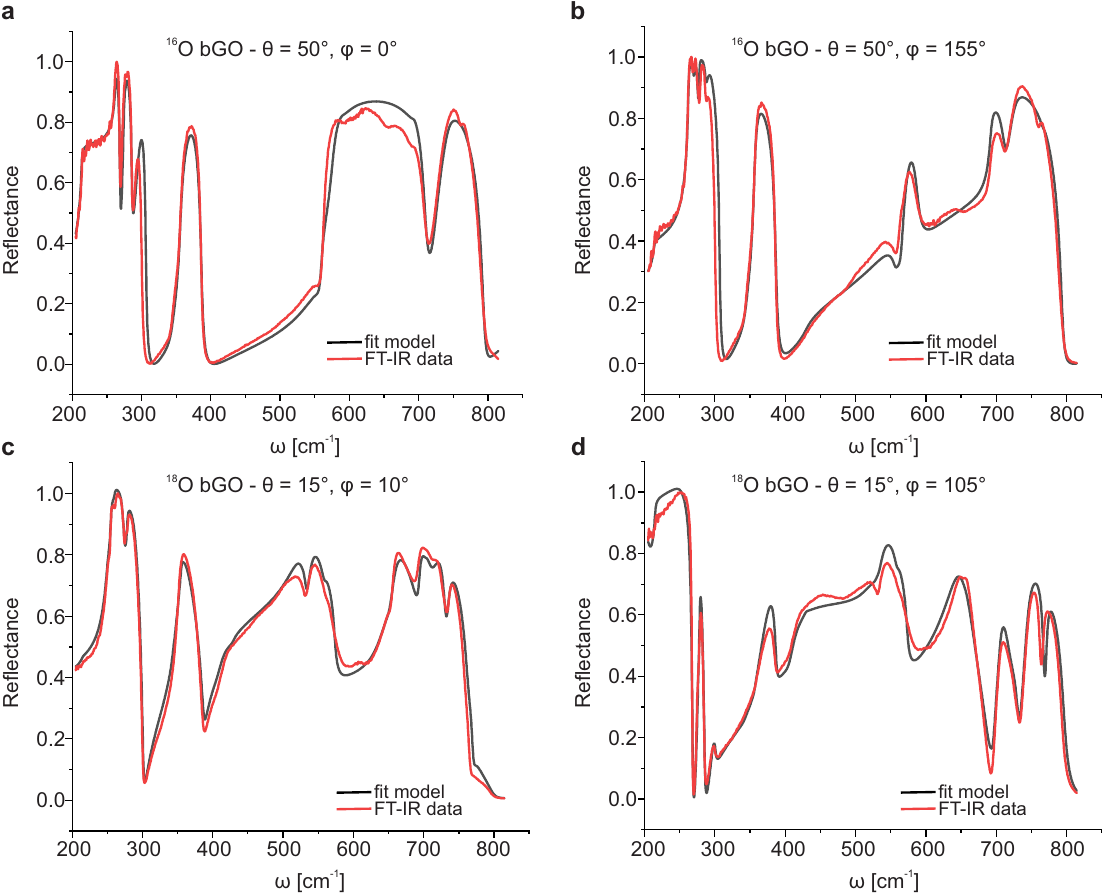}
    \caption{\textbf{Fitting procedure applied to derive the dielectric tensor from FT-IR measurements}. (a-b) Fitted (black lines) and raw experimental (red lines) plots for the \textsuperscript{16}O bGO bulk single crystal for azimuthal angles, $\phi$\,=\,0$^\circ$ (a) and $\phi$\,=\,155$^\circ$ (b). (c-d) Fitted (black lines) and raw experimental (red lines) plots for the \textsuperscript{18}O bGO epitaxial layer for azimuthal angles, $\phi$\,=\,10$^\circ$ (c) and $\phi$\,=\,105$^\circ$ (d). Note that the incidence angles $\theta$ chosen for the two isotopes are different, $\theta$\,=\,50$^\circ$ for \textsuperscript{16}O bGO and $\theta$\,=\,15$^\circ$ for \textsuperscript{18}O bGO.}
    \label{fig:figS5_2}
\end{figure}

\begin{figure}[ht!] 
\centering\includegraphics[width=1\textwidth]{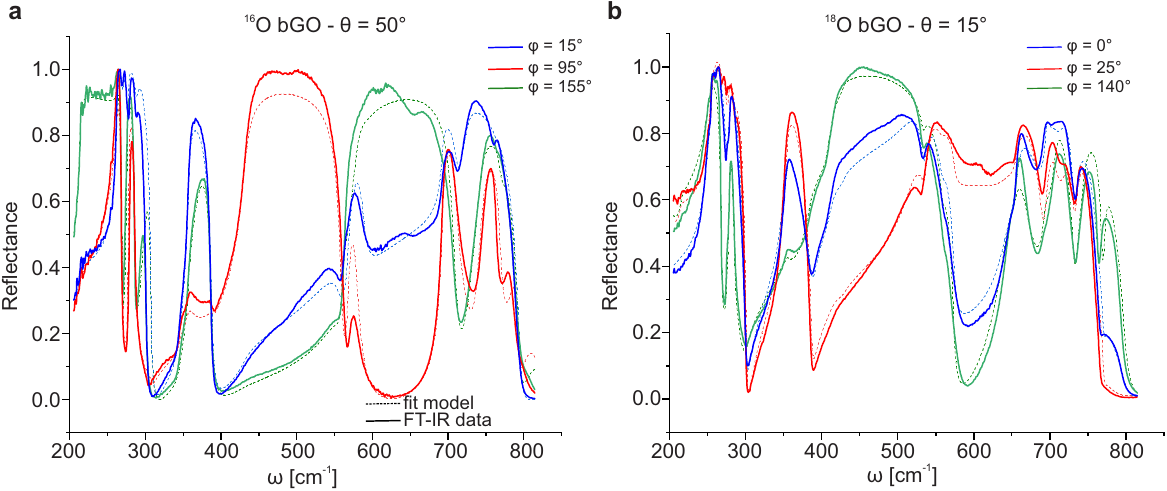}
    \caption{\textbf{Fitting procedure applied to derive the dielectric tensor from FT-IR measurements for various azimuthal angles}. (a) Fitted (dashed lines) and raw experimental (solid lines) plots for the \textsuperscript{16}O bGO bulk single crystal for azimuthal angles, $\phi$\,=\,15$^\circ$,\,95$^\circ$,\,155$^\circ$ at a fixed incidence angle, $\theta$\,=\,50$^\circ$. (b) Fitted (dashed lines) and raw experimental (solid lines) plots for the \textsuperscript{18}O bGO epitaxial layer for azimuthal angles, $\phi$\,=\,0$^\circ$,\,25$^\circ$,\,140$^\circ$ at a fixed incidence angle, $\theta$\,=\,15$^\circ$. }
    \label{fig:figS5_3}
\end{figure}
 
We could then proceed with the fitting for the \textsuperscript{16}O bGO isotope in \textit{WVASE}, starting from a single azimuthal angle, $\phi$, and adding additional azimuthal angles piecewise to achieve a systematic fit. The parameters for the \textsuperscript{18}O bGO were fitted following a similar approach, with the addition of a multilayer model accounting for the epitaxial structure - note that the sample consists of a \textsuperscript{16}O bGO substrate and a 1.2\,$\mu$m thick \textsuperscript{18}O bGO epitaxial layer - in the \textit{WVASE} software. While for the substrate layer of \textsuperscript{16}O bGO we made use of the previous fitting results, a new model was required for the epitaxial layer of \textsuperscript{18}O bGO. The initial parameters for the TO phonon frequencies in the case of the \textsuperscript{18}O bGO epitaxial layer were taken from \textit{ab initio} calculations, as shown in Table \,\ref{tab:epsparms}, while the remaining parameters were initially set to similar values as for the \textsuperscript{16}O bGO isotope, and they were all fitted. With the multilayer model complete, the first parameter optimized was the necessary Euler rotation, followed by the optimized parameters of the \textsuperscript{18}O layer. Finally, optimization is done with  \textsuperscript{16}O bGO parameters active along with  \textsuperscript{18}O bGO to ensure the contribution of generated reflectance spectra isn't only from the epi-layer. The fitting procedure uses the experimental data from \textsuperscript{18}O bGO measurements for the listed parameters, including the out-of-plane phonon modes (with A\textsubscript{u} symmetry).

\begin{figure}[ht!] 
\centering\includegraphics[width=1\textwidth]{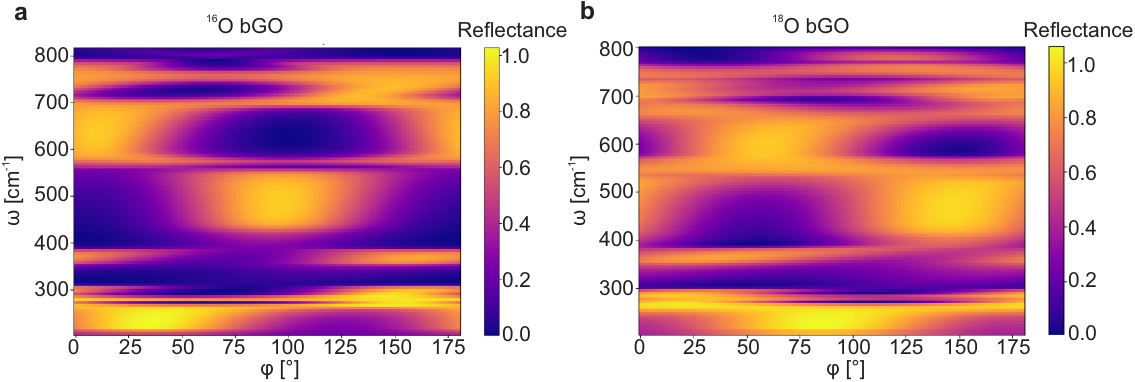}
    \caption{\textbf{FT-IR reflectance contour maps obtained from the \textit{WVASE} fitting results}. (a-b) FT-IR reflectance contour maps obtained from the \textit{WVASE} fitting results for the \textsuperscript{16}O bGO bulk single crystal (a) and the \textsuperscript{18}O bGO 1.2\,$\mu$m thick film homoepitaxially grown on the \textsuperscript{16}O bGO substrate (b).}
    \label{fig:figS5_4}
\end{figure}

Fig.\,\ref{fig:figS5_2} shows a comparison of the experimental raw data (red lines) with the fitting results (black lines) for both bGO isotopes at two different azimuthal orientations. This demonstrates that the fitting model accurately reproduces the reflectance features of the IR active modes for both materials. The fitting parameters have reasonable uncertainty values and capture the prominent features of the reststrahlen bands. However, although the fitting can capture these features, it has significant limitations when it comes to replicating the experimental data for the tiniest details in the reststrahlen bands. This can largely be attributed to systematic errors in the measurements. First, all data sets have a cutoff at 200\,cm\textsuperscript{-1}, that is, above the lowest frequency IR active mode (mode $\#$8) with B\textsubscript{u} symmetry, and rely on out-of-plane data from literature values only. Additionally, there are imperfections in the polarization state, angular spread and referencing inherent to reflection spectroscopy, all of which can introduce errors in measuring certain optical model parameters. Some of the systematic errors associated with these imperfections are discussed in the context of FT-IR microscopy in Ref.\,\cite{nandanwar2025determining}, but have not been systematically analyzed here. As such, while the model is sufficiently accurate for analysis of polariton data, it may not be fully representative of the true bGO optical constants. Azimuthal dispersion maps obtained from the fitting results are plotted in Figs.\,\ref{fig:figS5_4}.

\label{thepsilon}


\begin{table}[h!]
\fontsize{7pt}{7pt}\selectfont
\caption{\textbf{Oxygen isotope effect on in-plane transverse optical (TO) phonons in  bGO.} The experimental results were obtained from \textit{WVASE} fitting of polarized FT-IR reflectance measurements performed at various azimuthal angles (see SI Section \ref{expepsilon} \cite{schubert2019phonon}). All experimental values are reported with their fitting errors. The theoretical values are derived from \textit{ab initio} calculations, as shown in the Methods (main text).}
\label{tab:epsparms}
\centering
\begin{tabular}{l l l l l l l l l l l}
\toprule
 & \tabhead{B\textsubscript{u} mode} & Unit &  \tabhead{\#1} & \tabhead{\#2} & \tabhead{\#3} & \tabhead{\#4} & \tabhead{\#5} & \tabhead{\#6} & \tabhead{\#7} & \tabhead{\#8} \\
\hline\hline
Exp.  &  $\omega\textsubscript{TO,18}$ & cm\textsuperscript{-1} & 703.6 $\pm$ 0.3\ & 655.0 $\pm$ 0.4 & 538.1 $\pm$ 0.6 & 411.1 $\pm$ 0.5 & 352.9 $\pm$ 0.4 & 276.5 $\pm$ 0.4 & 254.9 $\pm$ 0.2 & 195 $\pm$ 2\\
Exp.  &  $\omega\textsubscript{TO,16}$ & cm\textsuperscript{-1} & 742.5 $\pm$ 0.2\ & 694.0 $\pm$ 0.1 & 567.2 $\pm$ 0.4 & 430.0 $\pm$ 0.3 & 356.2 $\pm$ 0.1 & 278.64 $\pm$ 0.08 & 260.5 $\pm$ 0.1 & 214.8 $\pm$ 0.1\\ 
Exp.  &  $\Delta\omega_{rel.}$ & \% & 5.2 & 5.6 & 5.1 & 4.4 & 0.9 & 0.8 & 2.1 & 9.2\\ 
\midrule
Th.  &   $\omega\textsubscript{TO,18}$ & cm\textsuperscript{-1} & 696.76 &  647.02 & 545.89 & 409.47 & 350.01 & 274.37 & 244.18 & 184.14\\
Th.  &   $\omega\textsubscript{TO,16}$ & cm\textsuperscript{-1} & 737.28 & 684.17 & 574.22 & 430.40 & 354.73& 277.30 & 252.68 & 193.74\\
Th.  &  $\Delta\omega_{rel.}$ & \% & 5.5 & 5.4 & 4.9 & 4.9 & 1.3 & 1.1 & 3.4 & 4.9\\ 
\hline\hline
Exp.  &  $ \lVert$\textbf{S}\textsubscript{18}$\rVert $ &   cm\textsuperscript{-1} & 281 $\pm$ 3 & 432 $\pm$ 3 & 650 $\pm$ 8 & 772 $\pm$ 6 & 326 $\pm$ 5 & 118 $\pm$ 6 & 437 $\pm$ 5 & 460 $\pm$ 10\\
Exp. &   $ \lVert$\textbf{S}\textsubscript{16}$\rVert $ &   cm\textsuperscript{-1} & 297 $\pm$ 2\ & 406 $\pm$ 2 & 810 $\pm$ 4 & 772 $\pm$ 3 & 349 $\pm$ 2 & 153 $\pm$ 2 & 445 $\pm$ 2 & 519 $\pm$ 2\\ 
\midrule
Th.  &   $ \lVert$\textbf{S}\textsubscript{18}$\rVert $  &   cm\textsuperscript{-1} & 436.54 & 578.37 & 1100.1 & 1059.5 & 401.22 & 90.09 & 571.25 & 639.42 \\
Th.  &  $ \lVert$\textbf{S}\textsubscript{16}$\rVert $  &   cm\textsuperscript{-1} &  436.28 & 604.5 & 1148.9 & 1103.1 & 454.56 & 161.09 & 569.52 & 674.99 \\
\hline\hline
Exp. &   $\delta$\textsubscript{18} & $^\circ$ & 53 $\pm$ 4 & 5 $\pm$ 3 & 108 $\pm$ 4 & 22 $\pm$ 3 & 146 $\pm$ 3 & 3 $\pm$ 4 & 165 $\pm$ 3 & 82 $\pm$ 4\\
Exp. &  $\delta$\textsubscript{16} & $^\circ$ & 48 $\pm$ 2 & 5 $\pm$ 2 & 108 $\pm$ 2 & 25 $\pm$ 2 & 145 $\pm$ 2 & 7 $\pm$ 2 & 159 $\pm$ 2 & 81 $\pm$ 2\\ 
\midrule
Th.  &   $\delta$\textsubscript{18} & $^\circ$ & 72.67 & 25.82 & 127.79 & 44.45
& 156.25 & 12.42
 & 182.03 & 105.26 \\
Th.  &  $\delta$\textsubscript{16} & $^\circ$ & 73.54 & 27.07 & 127.88 & 45.07 & 162.29 & 16.60 & 181.6 & 105.02 \\
\hline\hline
Exp.  &  $\gamma$\textsubscript{18} & cm\textsuperscript{-1} & 8.9 $\pm$ 0.3 & 13.5 $\pm$ 0.5 & 14.5 $\pm$ 0.6 & 16.2 $\pm$ 0.7 & 8.5 $\pm$ 0.5 & 6.0 $\pm$ 0.9 & 3.3 $\pm$ 0.3 & 1 $\pm$ 1\\
Exp.  & $\gamma$\textsubscript{16} & cm\textsuperscript{-1} & 13.2 $\pm$ 0.4 & 3.9 $\pm$ 0.1 & 17.6 $\pm$ 0.3 & 11.5 $\pm$ 0.3 & 4.3 $\pm$ 0.1 & 1.72 $\pm$ 0.08 & 2.3 $\pm$ 0.1 & 2.3 $\pm$ 0.1\\ 
\hline\hline
Th.  & $\gamma$\textsubscript{18} & cm\textsuperscript{-1} & 11.55 & 5.33 & 13.48 & 10.88 & 8.34 & 1.78 & 1.37 & 1.15\\
Th.  & $\gamma$\textsubscript{16} & cm\textsuperscript{-1} & 11.55 & 5.33 & 13.48 & 10.88 & 3.83 & 1.78 & 1.37 & 0.97\\ 
\bottomrule\\
\end{tabular}
\end{table}

\begin{table}[h!]
\fontsize{7pt}{7pt}\selectfont
\caption{\textbf{Electronic contributions to the dielectric tensor for the \textsuperscript{16}O and \textsuperscript{18}O bGO isotopes.}}
\label{tab:epsparms_inf}
\centering
\begin{tabular}{l l l l}
\toprule
  & \tabhead{$\varepsilon_{\infty,\text{xx}}$} & \tabhead{$\varepsilon_{\infty,\text{yy}}$} & \tabhead{$\varepsilon_{\infty,\text{zz}}$} \\
\hline\hline
Exp. \textsuperscript{18} & 3.69 $\pm$ 0.05 & 3.18 $\pm$ 0.06 & 4 $\pm$ 2 \\
Exp. \textsuperscript{16} &  3.78 $\pm$ 0.03 & 3.59 $\pm$ 0.03 & 4.042 $\pm$ 0.002 \\
\hline
Th. \textsuperscript{18} &  3.883  & 3.995 & 4.000 \\
Th. \textsuperscript{16} &  3.883 & 3.995 & 4.000 \\
\bottomrule\\
\end{tabular}
\end{table}

\section{Analytical derivation of the optical axis dispersion angle $\gamma$ and of the opening angle $\alpha$}


Fig.\,4 of the main text shows the optical axis and opening angle dispersion. The optical axis dispersion $\gamma(\omega)$ is measured as the angle between the coordinate system in which the permittivity tensor is measured (xyz) and the coordinate system that diagonalizes the real part of the permittivity tensor (mnz), as a function of frequency. At any given frequency, the basis vectors for the mnz coordinate system, in the xyz coordinates, are the eigenvectors $\mathbf{\hat e}$ of $\mathcal{R}(\boldsymbol{\varepsilon})$. The optical axis dispersion angle $\gamma$ is then the angle between the positive x-axis of the xyz frame and the eigenvector $\mathbf{\hat e}_{\text{m}}$. 

In the following, we will compare two different procedures for deriving the angle $\gamma$, in order to highlight the novelty of our approach with respect to previous studies \cite{passler2022hyperbolic,matson2023controlling}. The usual derivation method consists in applying a rotation to the original $\varepsilon$-tensor in Cartesian coordinates, $R(\gamma) \,\mathcal{R}(\varepsilon\textsubscript{xyz})\, R\textsuperscript{-1}(\gamma) = \mathcal{R}(\varepsilon\textsubscript{mnz})$. Setting the off-diagonal terms of $\varepsilon\textsubscript{mnz}$ to zero then leads to the equation:

\begin{equation}
\begin{aligned}
\label{gamma_rot}
    (\text{cos}^2(\gamma)-\text{sin}^2(\gamma))\mathcal{R}(\varepsilon_{\text{xy}})+\text{cos}(\gamma)\text{sin}(\gamma)(\mathcal{R}(\varepsilon_{\text{yy}})-\mathcal{R}(\varepsilon_{\text{xx}})) = \\
    = \text{cos}(2\gamma)\mathcal{R}(\varepsilon_{\text{xy}})+ \frac{1}{2}\text{sin}(2\gamma)(\mathcal{R}(\varepsilon_{\text{yy}})-\mathcal{R}(\varepsilon_{\text{xx}})) = 0
\end{aligned}
\end{equation}
Eq.\,\ref{gamma_rot} is solved by defining $\gamma$ as follows, 
\begin{equation}
\label{gamma}
\gamma = \frac{1}{2}\,\text{atan} \left(2\,\frac{\mathcal{R}{(\varepsilon_{xy})}}{\mathcal{R}{(\varepsilon_{xx})}-\mathcal{R}{(\varepsilon_{yy})}}\right).
\end{equation}
We notice here the existence of a 90$^\circ$ phase ambiguity in the definition of $\gamma$, due to the fact that eq.\,\ref{gamma_rot} may be equally solved by $\gamma' = \gamma +\frac{\pi}{2}$. When derived with this procedure, $\gamma$ is only defined within a restricted domain of existence, i.e., between $-\pi/4$ to $+\pi/4$, which does not correspond to the periodicity of the physical optical axes dispersion. Furthermore, another problem is that this method 
introduces discontinuities at the TO frequencies, which result in sudden phase jumps occurring as a consequence of the arctan function going from $-\pi/2$ to $+\pi/2$ when its argument oscillates between $-\infty$ to $+\infty$. 

\begin{figure}[ht!] 
\centering\includegraphics[width=0.7\textwidth]{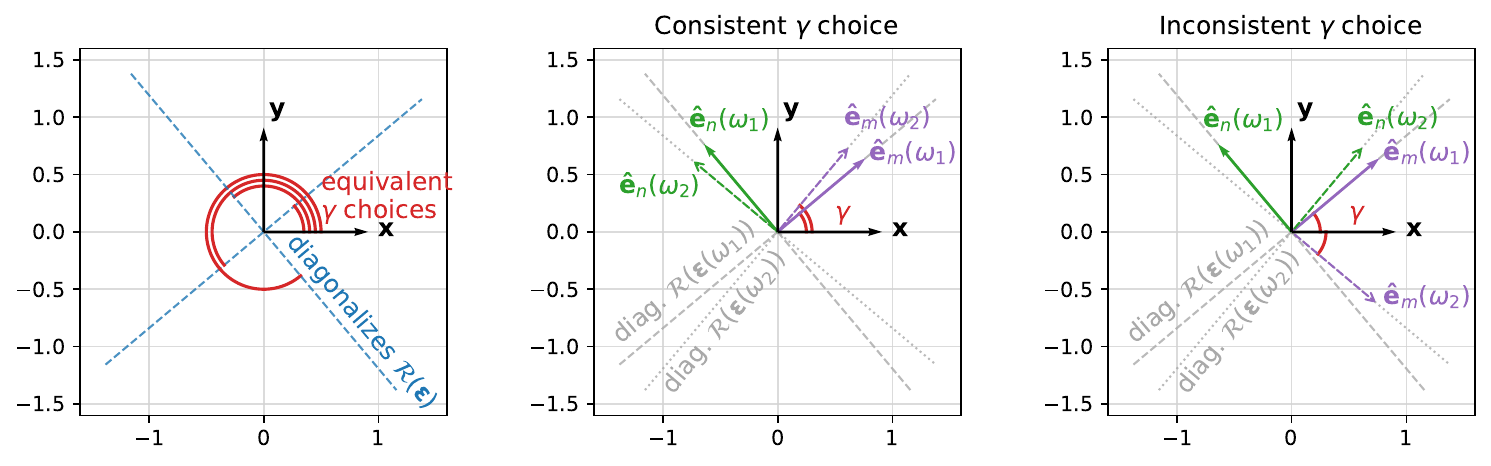}
    \caption{\textbf{The effect of inconsistent reference axis $\mathbf{\hat e}_m$ choice on the optical axis angle $\gamma$}. (a) Multiple values of $\gamma$, differing by multiples of 90 degrees, can describe the same mnz coordinate system that diagonalizes the permittivity tensor. As the mnz coordinate system rotates with frequency, the choice of the reference axis $\mathbf{\hat e}_m$ to which $\gamma$ is measured needs to be consistent. (b) A consistent choice of $\mathbf{\hat e}_{\text{m}}$ between frequencies $\omega_{\text{1}}$ and $\omega_{\text{2}}$. The change in $\gamma$ reflects the fact that the coordinate system that diagonalizes $\mathcal{R}(\boldsymbol{\varepsilon})$ has rotated by 10 degrees counter-clockwise. (c) An inconsistent choice of the mnz coordinate basis vectors $\mathbf{\hat e}_{\text{m}}$ and $\mathbf{\hat e}_{\text{n}}$ between frequencies $\omega_{\text{1}}$ and $\omega_{\text{2}}$. The change in gamma from 40 degrees to -40 degrees reflects the inconsistency of the axis/basis vector to which $\gamma$ has been measured, rather than the rotation of the optical axis by 80 degrees counter-clockwise.
    }
\label{fig:gamma_dispersion.consistent_ref_axis_coice_illustration}
\end{figure}
\begin{figure}[ht!] 
\centering\includegraphics[width=0.9\textwidth]{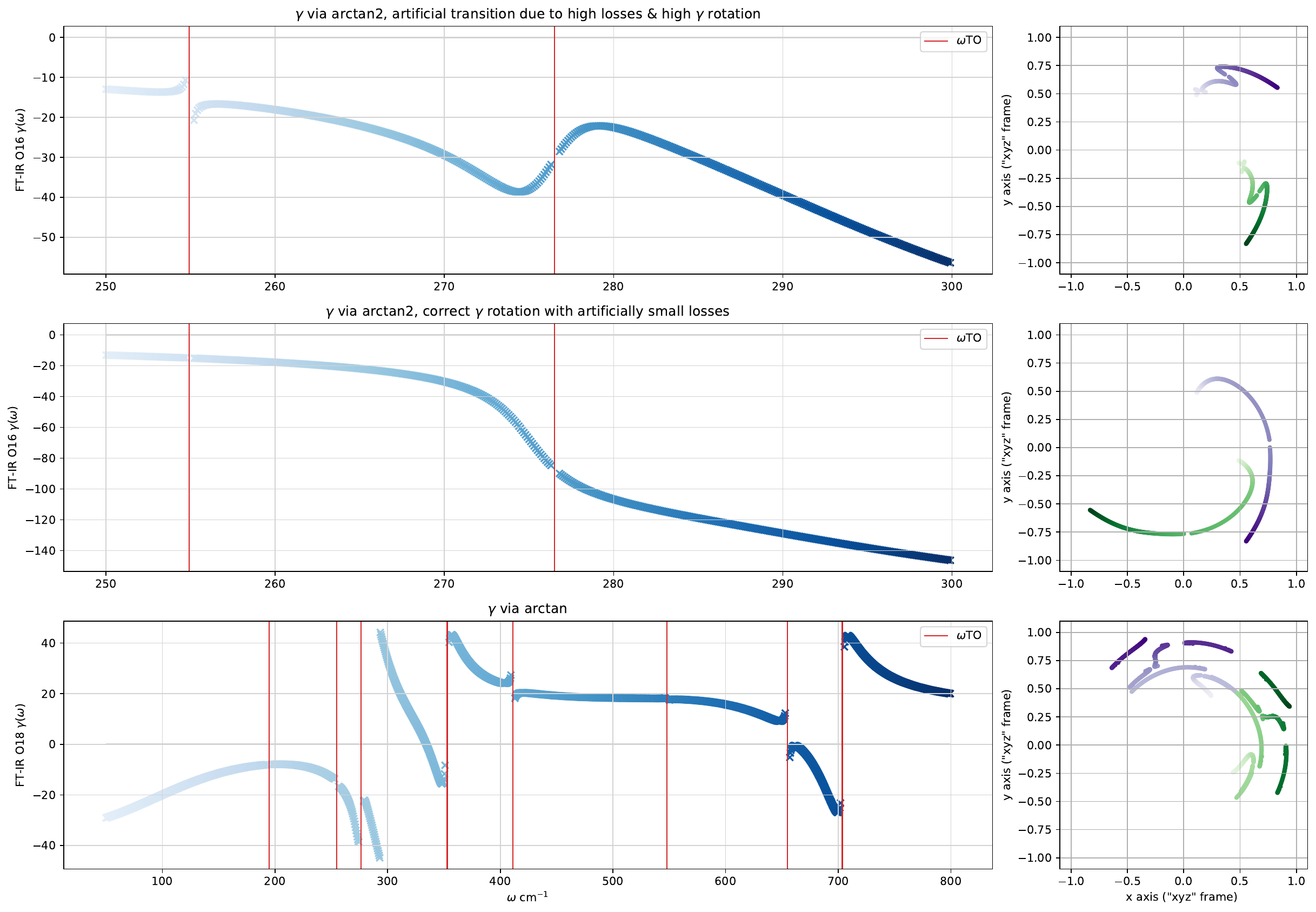}
  \caption{\textbf{Examples of challenges in computing $\gamma(\omega)$ in a consistent way across the full frequency range}. (a-b) $\gamma(\omega)$ and $\mathbf{\hat e}_{\text{m}}$ \& $\mathbf{\hat e}_{\text{n}}$ rotation with frequency, near mode \#6 $\omega_\textrm{TO}$ for the FT-IR O18 data. The permittivity was computed either with the experimentally measured Lorentzian broadening $\gamma_i$ (a) or with an arbitrarily small value (b). The fact that $\gamma$ shifts downwards (rotates clock-wise) through $\omega_\textrm{TO}$ in the permittivity model with artificially small broadening suggests that the wiggle in $\gamma$ computed with experimentally measured broadening is a failure of "eigenvector alignment" procedure of computing $\gamma$ rather than a physically meaningful rotation of the optical axis. In the rest of the analysis, the $\gamma(\omega)$ values to the right of $\omega_\textrm{TO}\approx 273\textrm{cm}^{-1}$ in panel (a) have been shifted down by 90 degrees to be consistent with the results in panel (b). (c) $\gamma(\omega)$ using eq.\,\ref{gamma}. Without further adjustment, $\gamma$ values are defined between $\pm$ 90 degrees, resulting in discontinuous rotation of $\mathbf{\hat e}$.}
\label{fig:gamma_dispersion.examples_of_challenges}
\end{figure}

The new procedure, developed in this work, tackles and solves these issues. Hereby, we present a thorough description of it. Instead of using eq.\,\ref{gamma}, we employ the equation that is given in the main text (eq.\,1), which relies on the eigenvectors of the dielectric tensor in cartesian coordinates, $\varepsilon_{\text{xyz}}$. For this method, the first step is to make a consistent choice about which of the two eigenvectors of $\mathcal{R}(\boldsymbol{\varepsilon})$ that span the monoclinic plane is chosen as the reference to which $\gamma$ is measured (see Fig \ref{fig:gamma_dispersion.consistent_ref_axis_coice_illustration}). Moreover, both $\mathbf{\hat e}_{\text{m}}$ and $-\mathbf{\hat e}_{\text{m}}$ are eigenvectors of $\mathcal{R}(\boldsymbol{\varepsilon})$, and a similarly consistent sign convention has to be kept across all frequencies at which $\gamma$ is computed. Otherwise, 90-degree discontinuities in $\gamma$ are introduced (e.g., Fig.~\ref{fig:gamma_dispersion.consistent_ref_axis_coice_illustration}\,c and Fig.~\ref{fig:gamma_dispersion.examples_of_challenges}\,c), which reflect an inconsistency in the choice of the basis vectors for the mnz coordinate system and not a physically meaningful rotation in the optical axis. The consistency is maintained by aligning the eigenvectors at frequency $\omega_{\text{i}}$ with those obtained at the previous frequency $\omega_{\text{i-1}}$. The alignment is done maximizing $\mathbf{\hat e}_{\text{m/n}} (\omega_{\text{i}}) \cdot \mathbf{\hat e}_{\text{m/n}} (\omega_{\text{i-1}})$ by permuting which eigenvector at $\omega_{\text{i}}$ is labeled $m$ and which - $n$ and by inverting them, as appropriate.

\begin{figure}[ht!] 
\centering\includegraphics[width=0.9\textwidth]{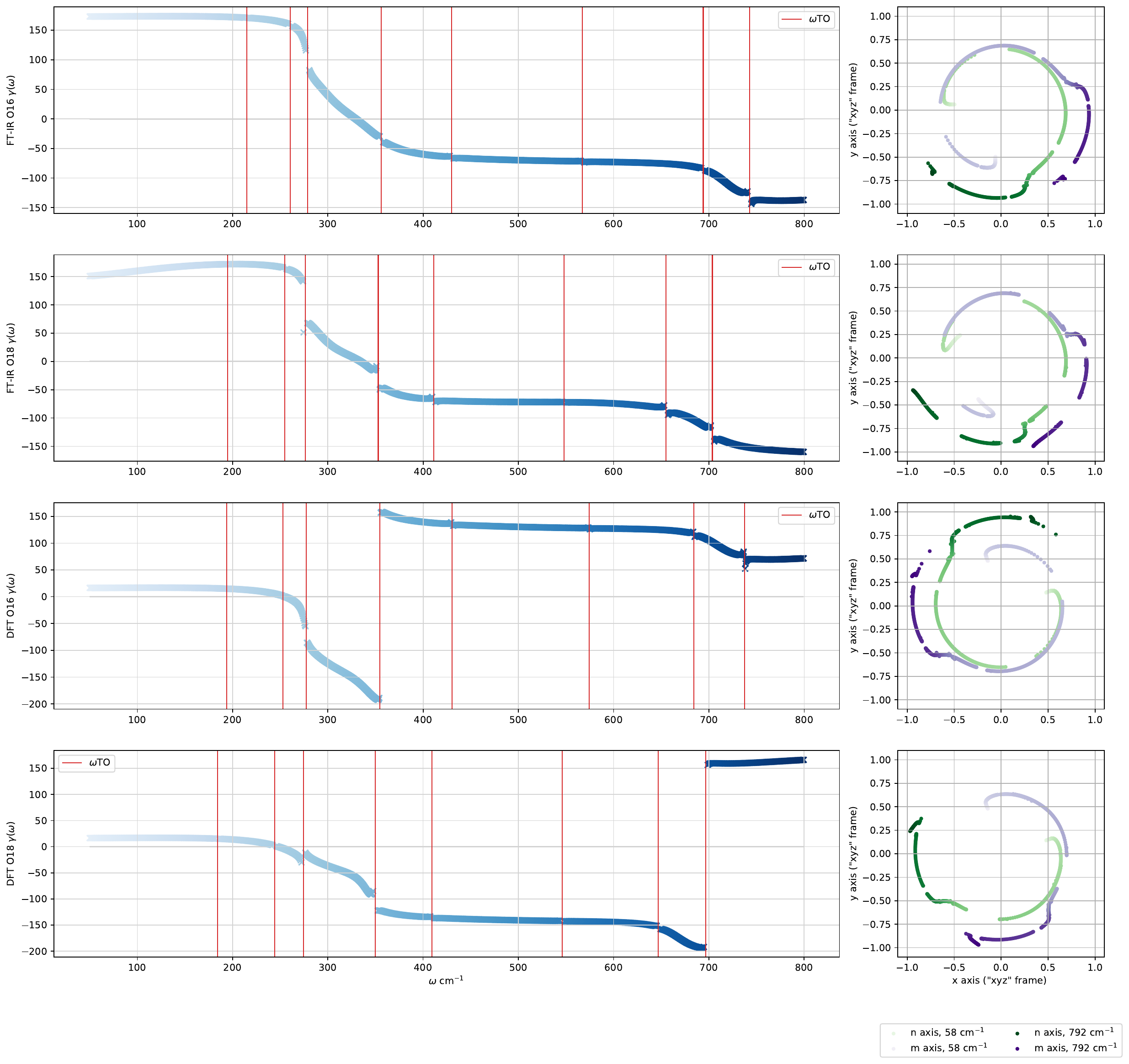}
    \caption{\textbf{Optical axis rotation with frequency for FT IR (a-b) and DFT (c-d) data, O16 (a, c) and O18 (b, d) bGO isotopologues.} 
     }
\label{fig:gamma_dispersion.all_data}
\end{figure}

The above-described "eigenvector alignment" procedure for determining $\gamma(\omega)$ is not foolproof, because the optical losses (Lorentzian damping constant $\gamma_{\text{i}}$ of eq.\,3, main text) smooth the transitions of $\boldsymbol{\varepsilon}$ across the resonance frequencies $\omega_{\textrm{TO}}$. 
This causes the $\mathbf{\hat e}_{\text{m}}$ to continuously rotate by $\sim$90 degrees across $\omega_{\textrm{TO}}$ resulting in a smooth, but artificial near-90-degree shifts in $\gamma(\omega)$. 
A convenient strategy is to exclude $\boldsymbol{\varepsilon}$ values about 1-2 cm$^{-1}$ of either side of $\omega_{\textrm{TO}}$. 
In most cases, the resulting eigenvectors of the last $\mathcal{R}(\boldsymbol{\varepsilon})$ value before and the first value after $\omega_{\textrm{TO}}$ are then aligned and chosen consistently. 
In the few remaining cases, the frequency range where the $\mathbf{\hat e}_{\text{m}}$ was chosen inconsistently with respect to those of the frequencies before, the resulting 90-degree shifts have to be accounted by on a case-by-case basis, for example, by shifting the corresponding $\gamma(\omega)$ region by an appropriate 90-degree-multiple shift. 
Since the origin of the smooth yet artificial $\mathbf{\hat e}_{\text{m}}$ rotation through $\omega_{\textrm{TO}}$ are the optical losses, the genuine cases of rapid $\gamma(\omega)$ rotation from the artificial ones may be distinguished by setting the Lorentzian broadening $\gamma_{\text{i}}$ to an arbitrary small value, see Fig. \ref{fig:gamma_dispersion.examples_of_challenges}. With this approach in place, $\gamma(\omega)$ is a 360-degree periodic and mostly rotates clockwise with frequency, with only a few rapid rotations around, e.g., at $\omega_\textrm{TO}\approx276 \textrm{cm}^{-1}$ - see Fig.~\ref{fig:gamma_dispersion.all_data}.

\begin{figure}[ht!] 
\centering\includegraphics[width=0.9\textwidth]{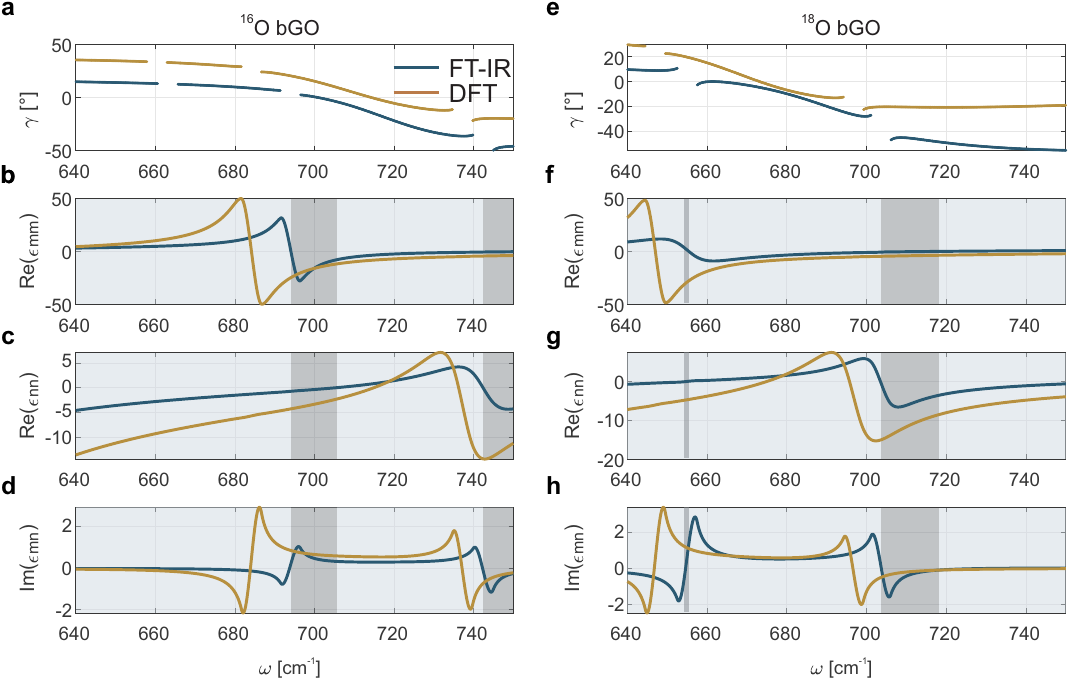}
    \caption{\textbf{Rotated $\varepsilon$-tensor components for the \textsuperscript{16}O and \textsuperscript{18}O bGO isotopes  extracted from FT-IR (blue) and DFT (orange) data}. (a,e) Frequency dependence of the optical axis dispersion angle $\gamma$ for the two bGO isotopes. (b,f,c,g) Real parts of the rotated $\varepsilon$-tensor diagonal elements for the two bGO isotopes. The shaded areas correspond to the spectral ranges where different types of phonon polaritons are supported: light blue indicates hyperbolic regions, whilst dark grey identifies elliptical polaritonic ranges. (d,h) Imaginary part of the off-diagonal element $\varepsilon\textsubscript{mn}$.}
    \label{fig:figS7_2}
\end{figure}

The angle $\gamma$ presented above can then be used to diagonalize $\varepsilon$\textsubscript{xyz}. The components of the $\varepsilon$-tensor in the rotated frame ($\varepsilon\textsubscript{mnz}$) are displayed in Fig.\,\ref{fig:figS7_2} for a restricted spectral range relevant for our experiment, i.e., 640-750\,cm\textsuperscript{-1}. The curves were derived from FT-IR and DFT data, see blue and orange curves, respectively. Note that the rotation applied can only diagonalize the real part of $\varepsilon\textsubscript{xyz}$. A full diagonalization of both the real and imaginary parts of the complex-valued dielectric tensor would require some other kind of transformation (involving non-hermitian operators, $\hat{O}\,\neq\,\hat{O}^{\dagger}$). Thus, in our case (since $\varepsilon$ is a complex-valued tensor), the imaginary parts of the off-diagonal elements do not disappear and are plotted in Fig.\,\ref{fig:figS7_2}\,d,h for both bGO isotopes. For a more in-depth explanation, see Ref.\,\cite{passler2022hyperbolic}. The light-blue and grey shaded areas in Figs.\,\ref{fig:figS7_2}\,b-d and f-h indicate the spectral ranges where different types of polaritons are supported: the two colors correspond to hyperbolic and elliptical regions, respectively. Note that the polariton bands were classified based on the zero crossings calculated from the FT-IR permittivity data. In the left panels relative to the \textsuperscript{16}O bGO isotope, we observe two elliptical regions of similar widths, i.e., spectral ranges where the real parts of both in-plane diagonal elements of the $\varepsilon\textsubscript{mnz}$-tensor are negative. 
The first elliptical region for the \textsuperscript{18}O bGO isotope, shown in Fig.\,\ref{fig:figS7_2}\,g at $\omega$\,$\approx$\,655\,cm\textsuperscript{-1}, is much narrower, which is due to a combination of two main factors: 1) the large damping constant $\gamma$\textsubscript{18} of the B\textsubscript{u} mode $\#$2 for the \textsuperscript{18}O isotope (by a factor of 3.5  larger than the corresponding value for the \textsuperscript{16}O isotope); 2) the smaller damping constant $\gamma$\textsubscript{18} of the B\textsubscript{u} mode $\#$1 compared to the corresponding value for the \textsuperscript{16}O isotope.

The angle $\alpha$ is derived geometrically from the ratio of the in-plane k-vector components in the newly defined reference frame, i.e., k\textsubscript{mm} and k\textsubscript{nn}. Note that, for in-plane hyperbolic biaxial crystals, this ratio is related to the permittivity components by the equation $\frac{k\textsubscript{mm}}{k\textsubscript{nn}} = \sqrt{-\frac{\mathcal{R}(\varepsilon\textsubscript{nn})}{\mathcal{R}(\varepsilon\textsubscript{mm})}}$, as demonstrated in Ref.\,\cite{alvarez2019analytical}. To summarize, $\alpha$ is defined by the following expression:

\begin{equation}
\label{alpha}
\alpha = 2 \,\text{atan}\left(\frac{k\textsubscript{mm}}{k\textsubscript{nn}}\right) = 2 \,\text{atan}\left(\sqrt{-\frac{\mathcal{R}(\varepsilon\textsubscript{nn})}{\mathcal{R}(\varepsilon\textsubscript{mm})}}\right).
\end{equation}

\section{Error estimation for s-SNOM and FT-IR data}

This section is aimed at explaining the error estimation for both s-SNOM and far-field IR-reflectance data plotted in Fig.\,4 of the main text. The vertical error bars have been calculated from the uncertainties on the weighted average of the long and short ray's angles ($\theta\textsubscript{lr,1/2}$, $\theta\textsubscript{sr}$) as follows:
\begin{equation}
    (\sigma_{\bar{x}})_{wtd} = \sqrt{\frac{\sum_{i=1}^{n}(w_i(x_i-\bar{x}_{wtd})^2)}{\sum_{i=1}^{n}w_i} \frac{1}{n_{eff}-1}}
\end{equation}
where $n_{eff}$ is the effective number of degrees of freedom, $n_{eff} =\frac{(\sum_{i=1}^{n}w_i)^2}{\sum_{i=1}^{n}(w_i^2)}$. In this way, the standard deviation from the weighted average is taken into account and the effective number of measurements depends on their weights. The horizontal error bars have been set to a length of 4\,cm\textsuperscript{-1}, assuming a 0.5$\%$ FWHM of the incident FEL frequency.

We also included the min/max estimate for oscillator strength and damping (1. highest oscillator strength and lowest damping, 2. lowest oscillator strength and highest damping) from the \textit{WVASE} fits to the IR-reflectance maps, in addition to a $\pm$ 0.5 cm\textsuperscript{-1} horizontal uncertainty to account for the errors on the TO phonon frequencies. 

\clearpage

\section{Crystal orientation}

\begin{figure}[h] 
\centering\includegraphics[width=0.9\textwidth]{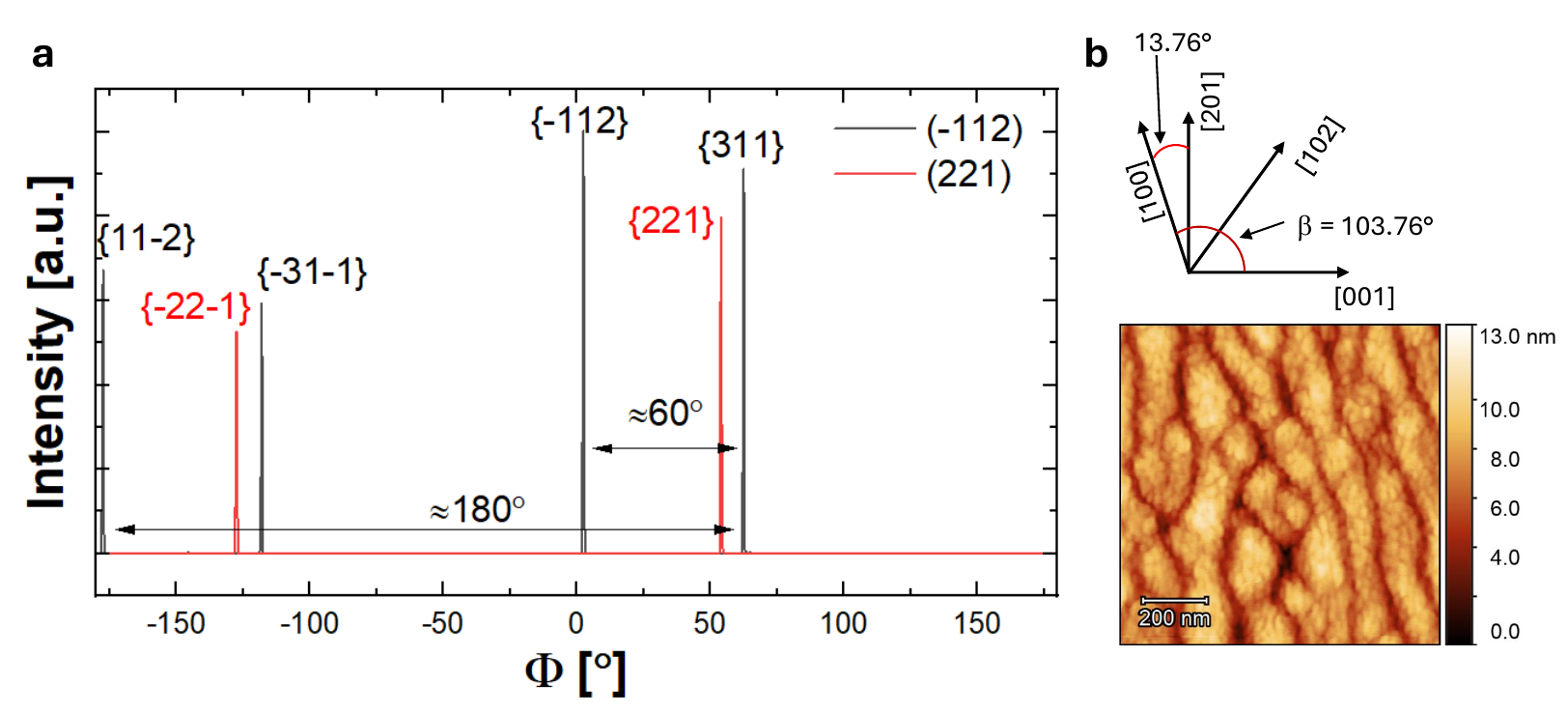}
    \caption{\textbf{Determination of the bGO In-plane orientations.} (a) $\Phi$-scans for the (-112) and (221) planes to determine the [001] in-plane direction of the \textsuperscript{18}bGO sample. (b) AFM image of the bGO with the reference crystallographic directions as determined by XRD.}
    \label{fig:figS9_1}
\end{figure}

X-ray diffraction (XRD) on a four-circle diffractometer (Malvern Panalytical X’Pert MRD) was performed on the \textsuperscript{18}bGO to determine its in-plane crystallographic orientations. To probe the in-plane component for the investigated (010) homolayer, $\Phi$-scans were performed for the (-112) and (221) planes, which give information for the [001] and [102] orientations, respectively. The spacing of the reflexes was confirmed by the stereographic projection generated using \textit{WinWulff} and labeled as reported in Fig.\,\ref{fig:figS9_1}(a). It is important to note that due to the two-fold symmetry of the (010) surface, only the [001] direction can be highlighted but not its absolute value (+ or -c), regardless, for the same reason, this does not play a relevant role for the current investigation. To confirm the crystallographic orientation extracted from the XRD measurements, additional AFM measurements were performed (Fig.\,\ref{fig:figS9_1}(b)); in fact, (010) oriented homoepitaxial layers usually show peculiar features on its growth surface, i.e.,  the elongated features which are mostly orthogonal to the [001], trace of the anisotropic surface ad-atom diffusion during growth \cite{Mazzolini_2020,AhmadiDisilane}.

\section{Penetration depth estimation of the polariton evanescent wave inside the \textsuperscript{18}O bGO epitaxial film}

To estimate the penetration depth of the SPhPs in our sample, we use the following formula:

\begin{equation}
d_{\pm} = \frac{1}{\kappa_{\pm}}
\end{equation}
where $\kappa_{\pm}$ is $i k_z$, i.e., the $z$-component of the $k$-vector. In the presence of evanescence, the decay constants $\kappa_{\pm}$ are always real and positive. The notation $\pm$ indicates the specific evanescent wave under consideration, with each wave decaying into one of the two half-spaces, filled with bGO (+) and air (-), respectively. By applying the boundary conditions on the wave vector components parallel to the interface, we utilize the dispersion relation of bulk phonon polaritons in isotropic materials to derive the following expression for $\kappa_{\pm}$:

\begin{equation}
\kappa_{\pm} = k_0\sqrt{k_{x,0}^2-\varepsilon_{\pm}}
\end{equation}
Considering incident light with a wavenumber, $\omega_0\,=\, 700$\,cm\textsuperscript{-1}, we find the corresponding wavevector magnitude to be $k_0=2\pi\,\omega_0=0.44\,$\textmu m\textsuperscript{-1}. On the bGO side, we have a large negative $\varepsilon_{-}$, due to the material being in the reststrahlen band, specifically, $\varepsilon_{zz} \approx -4 $ along the $z$-direction. This results in a penetration depth on the bGO side of approximately $d_{\text{bGO}} = 0.23\,$\textmu m. In contrast, the skin depth, which is relevant for electric fields, is calculated to be $0.46\,$\textmu m. This difference of a factor of 2 between the two values arises because the penetration depth is calculated for intensities, while the skin depth pertains to electric fields. When considering an Au disc with a thickness of $4\,$\textmu m, the skin depth into the bGO half-space is estimated to be $0.84\,$\textmu m, assuming $k_{x,0} \approx 5$.

\bibliographystyle{plain}
\bibliography{bibliography}